**This is a preprint version from prior to peer review.**

An updated, peer reviewed version of the article has been published as "Artificial intelligence in digital pathology: a systematic review and meta-analysis of diagnostic test accuracy" at NPJ Digital Medicine

This is available open access here:
https://www.nature.com/articles/s41746-024-01106-8#citeas

**TITLE:** Artificial intelligence in digital pathology: a diagnostic test accuracy systematic review and meta-analysis

**AUTHORS:** Clare McGenity[1,2], Emily L Clarke[1,2], Charlotte Jennings[1,2], Gillian Matthews[2], Caroline Cartlidge[1], Henschel Freduah-Agyemang[1], Deborah D Stocken[1], Darren Treanor[1,2,3,4]

**AFFILIATIONS:**
[1] University of Leeds, Leeds, UK
[2] Leeds Teaching Hospitals NHS Trust, Leeds, UK
[3] Department of Clinical Pathology and Department of Clinical and Experimental Medicine, Linköping University, Linköping, Sweden
[4] Centre for Medical Image Science and Visualization (CMIV), Linköping University, Linköping, Sweden

**CORRESPONDING AUTHOR:**
Dr Clare McGenity
Email: c.m.mcgenity@leeds.ac.uk

**ABSTRACT:**

**Background**
Ensuring diagnostic performance of artificial intelligence (AI) before introduction into clinical practice is key to safe and successful adoption of this technology. Growing numbers of studies using AI for digital pathology have been reported over recent years. The aim of this work is to examine the diagnostic accuracy of AI in digital pathology images for any disease.

**Methods**
This systematic review and meta-analysis included diagnostic accuracy studies using any type of artificial intelligence applied to whole slide images (WSIs) for any disease. The reference standard was diagnosis by histopathological assessment and / or immunohistochemistry. Searches were conducted in PubMed, EMBASE and CENTRAL in June 2022. Risk of bias and concerns of applicability were assessed using the QUADAS-2 tool. Data extraction was conducted by two investigators and meta-analysis was performed using a bivariate random effects model.

**Results**
Of 2976 identified studies, 100 were included in the review and 48 in the meta-analysis. These studies were from a broad range of countries, including over 152,000 whole slide images (WSIs) and representing many diseases, which were predominantly cancers but also other conditions. These studies reported a mean sensitivity of 96.3% (CI 94.1-97.7) and mean specificity of 93.3% (CI 90.5-95.4) for AI in WSIs. There was heterogeneity in study design and 99% of studies identified for inclusion had at least one area at high or unclear risk of bias.

**Conclusions**
This review provides an overview of AI performance in whole slide imaging. Studies had variability in dataset sizes, dataset descriptions, unit of analysis, study design and available performance data. Details around the selection of cases, division of data for model development and validation and raw performance data were frequently ambiguous or missing. Overall, AI is reported as having high diagnostic accuracy in the reported areas but requires more rigorous evaluation of its performance.

## INTRODUCTION:

Following recent prominent discoveries in deep learning techniques, wider AI applications have emerged for many sectors, including in healthcare.[1-3] Pathology AI is of broad importance in areas across medicine, with implications not only in diagnostics, but in cancer research, clinical trials and AI-enabled therapeutic targeting.[4] Application of AI to an array of diagnostic tasks using whole slide images (WSIs) has rapidly expanded in recent years.[5-8] Successes in AI for digital pathology can be found for many disease types, but particularly in examples applied to cancer.[4,9-11] An important early study in 2017 by Bejnordi et al. described 32 AI models developed for breast cancer detection in lymph nodes through the CAMELYON16 grand challenge. The best model achieved an area under the curve (AUC) of 0.994 (95% CI 0.983-0.999), demonstrating similar performance to the human in this controlled environment.[12] A study by Lu et al. in 2021 trained AI to predict tumour origin in cases of cancer of unknown primary (CUP).[13] Their model achieved an AUC of 0.8 and 0.93 for top-1 and top-3 tumour accuracies respectively on an external test set. AI has also been applied to making predictions, such as determining the 5-year survival in colorectal cancer patients and the mutation status across multiple tumour types.[14,15]

Several reviews have examined the performance of AI in subspecialties of pathology. In 2020, Thakur et al. identified 30 studies of colorectal cancer for review with some demonstrating high diagnostic accuracy, although the overall scale of studies was small and limited in their clinical application.[16] Similarly in breast cancer, Krithiga et al. examined studies where image analysis techniques were used to detect, segment and classify disease, with reported accuracies ranging from 77 to 98%.[17] Other reviews have examined applications in liver pathology, skin pathology and kidney pathology with evidence of high diagnostic accuracy from some AI models.[18-20] Additionally, Rodriguez et al. performed a broader review of AI applied to WSIs and identified 26 studies for inclusion with a focus on slide level diagnosis.[21] They found substantial heterogeneity in the way performance metrics were presented and limitations in the ground truth used within studies. However, their study did not address other units of analysis and no meta-analysis was performed. Therefore, the present study is the first systematic review and meta-analysis to address the diagnostic accuracy of AI across all disease areas in digital pathology, and includes studies with multiple units of analysis.

Despite the many developments in pathology AI, examples of routine clinical use of these technologies remain rare and there are concerns around the performance, evidence quality and risk of bias for medical AI studies in general.[22-24] Although, in the face of an increasing pathology workforce crisis, the prospect of tools that can assist and automate tasks is appealing.[25,26] Challenging workflows and long waiting lists mean that substantial patient benefit could be realised if AI was successfully harnessed to assist in the pathology laboratory.

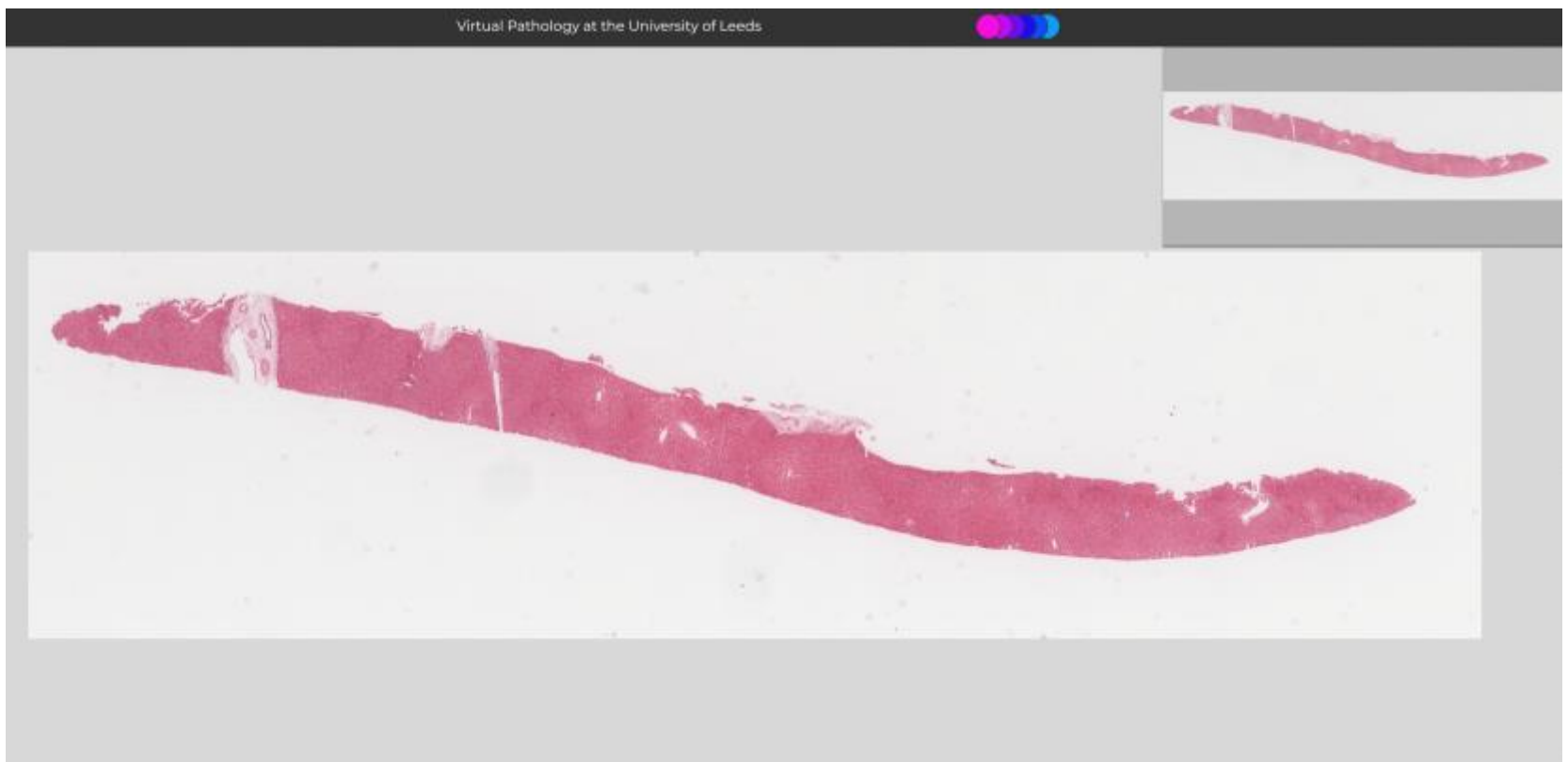

***Figure 1*** *– Example whole slide image (WSI) of a liver biopsy specimen at low magnification. These are high resolution digital pathology images viewed by a pathologist on a computer to make a diagnostic assessment. Image courtesy of www.virtualpathology.leeds.ac.uk*[27]

This systematic review provides an overview of performance of diagnostic tools across histopathology. The objective of this review was to determine the diagnostic test accuracy of artificial intelligence solutions applied to WSIs to diagnose disease.

**METHODS:**

This systematic review and meta-analysis was conducted in accordance with the guidelines for the “Preferred Reporting Items for Systematic Reviews and Meta-Analyses” extension for diagnostic accuracy studies (PRISMA-DTA).[28] The protocol for this review is available https://www.crd.york.ac.uk/prospero/display_record.php?ID=CRD42022341864 (Registration: CRD42022341864).

Eligibility Criteria

Studies reporting the diagnostic accuracy of AI models applied to WSIs for any disease diagnosed through histopathological (surgical pathology) assessment and / or immunohistochemistry were sought. The primary outcome was the diagnostic accuracy of AI tools in detecting disease or classifying subtypes of disease. The index test was any AI model applied to WSIs. The reference standard was any diagnostic histopathological interpretation by a pathologist and / or immunohistochemistry.

Studies were excluded where the outcome was a prediction of patient outcomes, treatment response, molecular status, whilst having no detection or classification of disease. Studies of cytology, autopsy and forensics cases were excluded. Studies grading, staging or scoring disease, but without results for detection of disease or classification of disease subtypes were also excluded. Studies examining modalities other than whole slide imaging or studies where WSIs were mixed with other imaging formats were also excluded.

Data sources and search strategy

Electronic searches of PubMed, EMBASE and CENTRAL were performed from inception to 20th June 2022. Searches were restricted to English language and human studies. There were no restrictions on the date of publication. The full search strategy is available in the supplementary materials. Citation checking was also conducted.

Study selection

Two investigators (C.M. and H.F.A.) independently screened titles and abstracts against a predefined algorithm to select studies for full text review. The screening tool is available in the supplementary materials. Disagreement regarding study inclusion was resolved by discussion with a third investigator (D.T.). Full text articles were reviewed by two investigators (C.M. and E.L.C.) to determine studies for final inclusion.

Data extraction and quality assessment

Data collection for each study was performed independently by two reviewers using a predefined electronic data extraction spreadsheet. Every study was reviewed by the first investigator (C.M.) and a team of four investigators were used for second independent review (E.L.C. / C.J. / G.M. / C.C.). Data extraction obtained the study demographics; disease examined; pathological subspecialty; type of AI; type of reference standard; datasets details; split into train / validate / test sets and test statistics to construct 2x2 tables of the number of true-positives (TP), false positives (FP), false negatives (FN) and true negatives (TN). An indication of best performance with any diagnostic accuracy metric provided was recorded for all studies. Corresponding authors of the primary research were contacted to obtain missing performance data for inclusion in the meta-analysis.

At the time of writing, the QUADAS-AI tool was still in development and so could not be utilised.[29] Therefore, a tailored QUADAS-2 tool was used to assess the risk of bias and any applicability concerns for the included studies.[30,31] Further details of the quality assessment process can be found in the supplementary materials.

Statistical analysis

Data analysis was performed using MetaDTA: Diagnostic Test Accuracy Meta-Analysis v2.01 Shiny App to generate forest plots, summary receiver operating characteristic (SROC) plots and summary sensitivities and specificities, using a bivariate random effects model.[32,33] If available, 2x2 tables were used to include studies in the meta-analysis to provide an indication of diagnostic accuracy demonstrated in the study. Where unavailable, this data was requested from authors or calculated from other metrics provided. Where only multiclass data was available, this was combined into a 2x2 format, unless negative results categories were unavailable (e.g. for multiple comparisons between disease types only). Sensitivity and specificity were examined overall and in the largest pathological subspecialty groups to compare diagnostic accuracy among these studies.

## RESULTS

Study selection

Searches identified 2976 abstracts, of which 1666 were screened after duplicates were removed. 296 full text papers were reviewed for potential inclusion. 100 studies met the full inclusion criteria for inclusion in the review, with 48 studies included in the full meta-analysis (***Figure 2***).

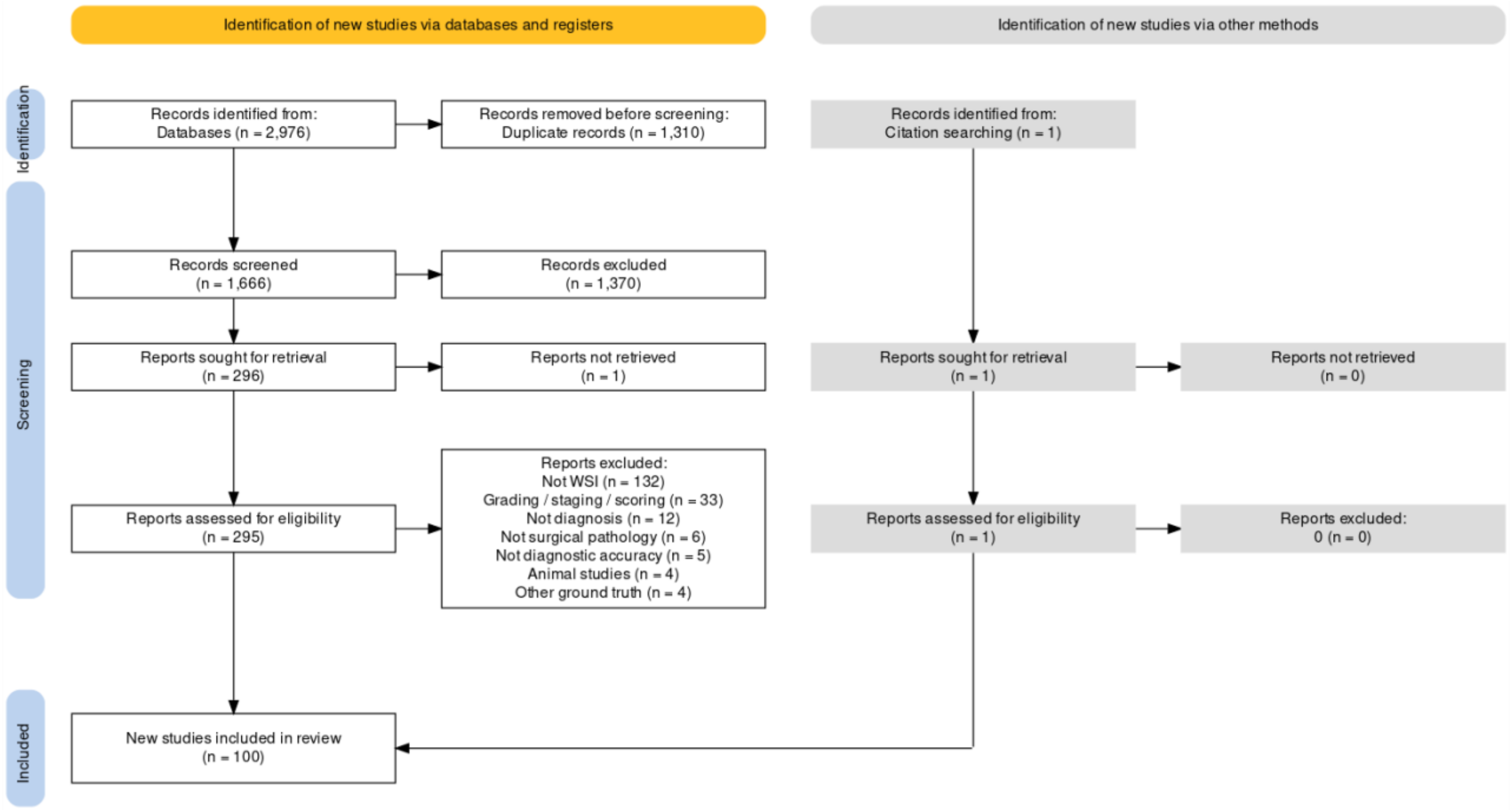

***Figure 2*** *– Study selection flow diagram. Generated using PRISMA2020 at* *https://estech.shinyapps.io/prisma_flowdiagram/* [34]

Study characteristics

Study characteristics are presented by pathological subspecialty for all 100 studies identified for inclusion in ***Tables 1-7***. Studies from Europe, Asia, Africa, North America, South America and Australia / Oceania were all represented within the review, with the largest numbers of studies coming from the USA and China. Total numbers of images used across the datasets equated to over 152,000 WSIs. Further details, including funding sources for the studies can be found in the supplementary materials. ***Table 1*** and ***Table 2*** show characteristics for breast pathology and cardiothoracic pathology studies respectively. ***Table 3*** and ***Table 4*** are characteristics for dermatopathology and hepatobiliary pathology studies respectively. ***Table 5*** and ***Table 6*** have characteristics for gastrointestinal and urological pathology studies respectively. Finally, ***Table 7*** outlines characteristics for studies with multiple pathologies examined together and for other pathologies such as gynaepathology, haematopathology, head and neck pathology, neuropathology, paediatric pathology, bone pathology and soft tissue pathology.

Risk of bias and applicability

The risk of bias and applicability assessment using the tailored QUADAS-2 tool demonstrated that the majority of papers were either at high risk or unclear risk of bias in three out of the four domains (***Figure 3***). The full breakdown of individual paper scores can be found in the supplementary materials. Of the 100 studies included in the systematic review, 99% demonstrated at least one area at high or unclear risk of bias, with many having multiple components at risk.

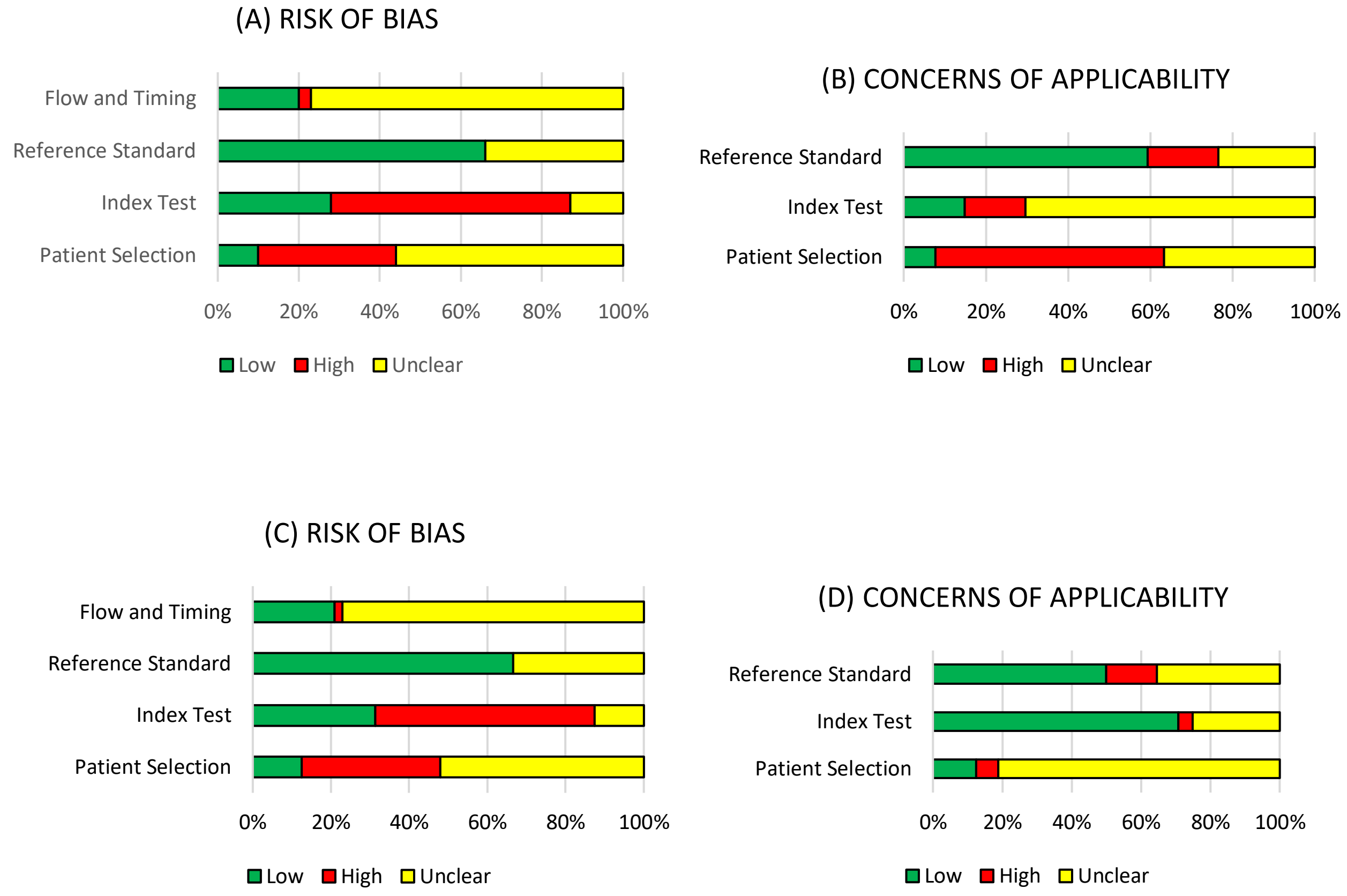

***Figure 3*** *– Risk of bias and concerns of applicability in summary percentages for studies included in the review. (A) & (B): Summaries for all 100 papers included in the review. (C) & (D): Summaries for 48 papers included in the meta-analysis.*

Of the 48 studies included in the meta-analysis (***Figure 3C*** and ***Figure 3D)***, 42 of 48 studies were either at high or unclear risk of bias for patient selection and 33 of 48 studies were at high or unclear risk of bias concerning the index test. The most common reasons for this included cases not being selected randomly or consecutively, or the selection method being unclear, the absence of external validation of the study's findings and a lack of clarity on whether training and testing data were mixed. 16 of 48 studies were unclear in terms of their risk of bias for the reference standard, but no studies were considered high risk in this domain. For flow and timing, one study was at high risk but 37 of 48 studies were at unclear risk of bias.

There were concerns of applicability for many papers included in the meta-analysis with 42 of 48 studies with either unclear or high concerns for applicability in the patient selection, 14 of 48 studies with unclear or high concern for the index test and 24 of 48 studies with unclear or high concern for the reference standard. Examples for this included ambiguity around the selection of cases and the risk of excluding subgroups, and limited or no details given around the diagnostic criteria and pathologist involvement when describing the ground truth.

Synthesis of results

100 studies were identified for inclusion in this systematic review. Included study size varied greatly from 4 WSIs to nearly 30,000 WSIs. Data on a WSI level was frequently unavailable for numbers used in test sets, but where it was reported this ranged from 10 WSI to nearly 14,000 WSIs, with a mean of 822 WSIs. The majority of studies had small datasets and just a few studies contained comparatively large datasets of thousands or tens of thousands of WSIs. Of included studies, 48 had data that could be meta-analysed. Two of the studies in the meta-analysis had available data for two different disease types,[35,36] meaning a total of 50 assessments included in the meta-analysis. ***Figure 4*** shows the forest plots for sensitivity of any AI solution applied to whole slide images. Overall, there was high diagnostic accuracy across studies and disease types. The mean sensitivity across all studies was 96.3% (CI 94.1-97.7) and mean specificity was 93.3% (CI 90.5-95.4), as shown in ***Figure 5***.

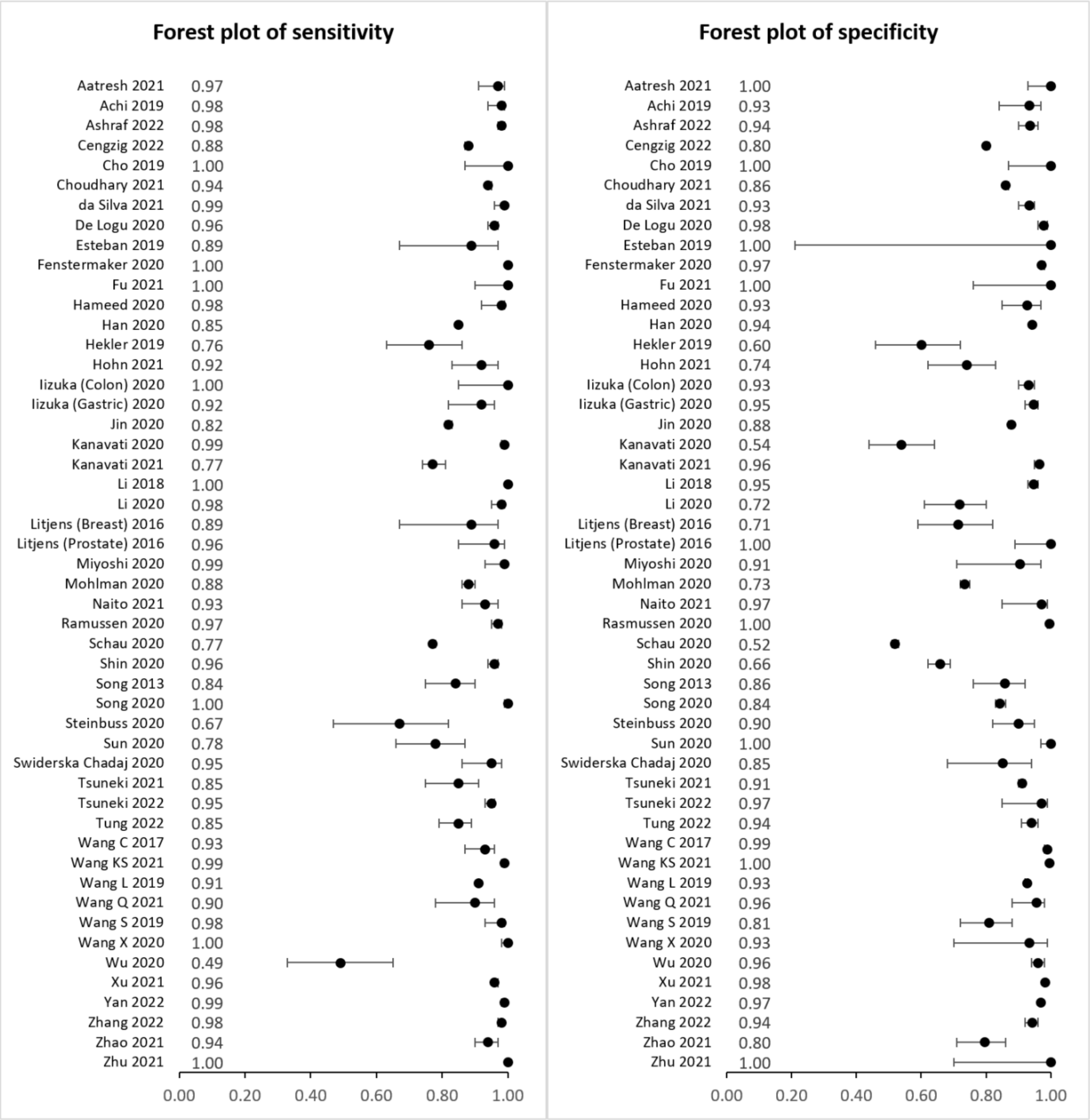

***Figure 4*** – Forest plots for sensitivity and specificity in studies of all pathologies with 95% confidence intervals. Data and error bar values used in these plots were generated by MetaDTA: Diagnostic Test Accuracy Meta-Analysis v2.01 Shiny App https://crsu.shinyapps.io/dta_ma/ and the data can be found in the supplementary materials.[32,33]

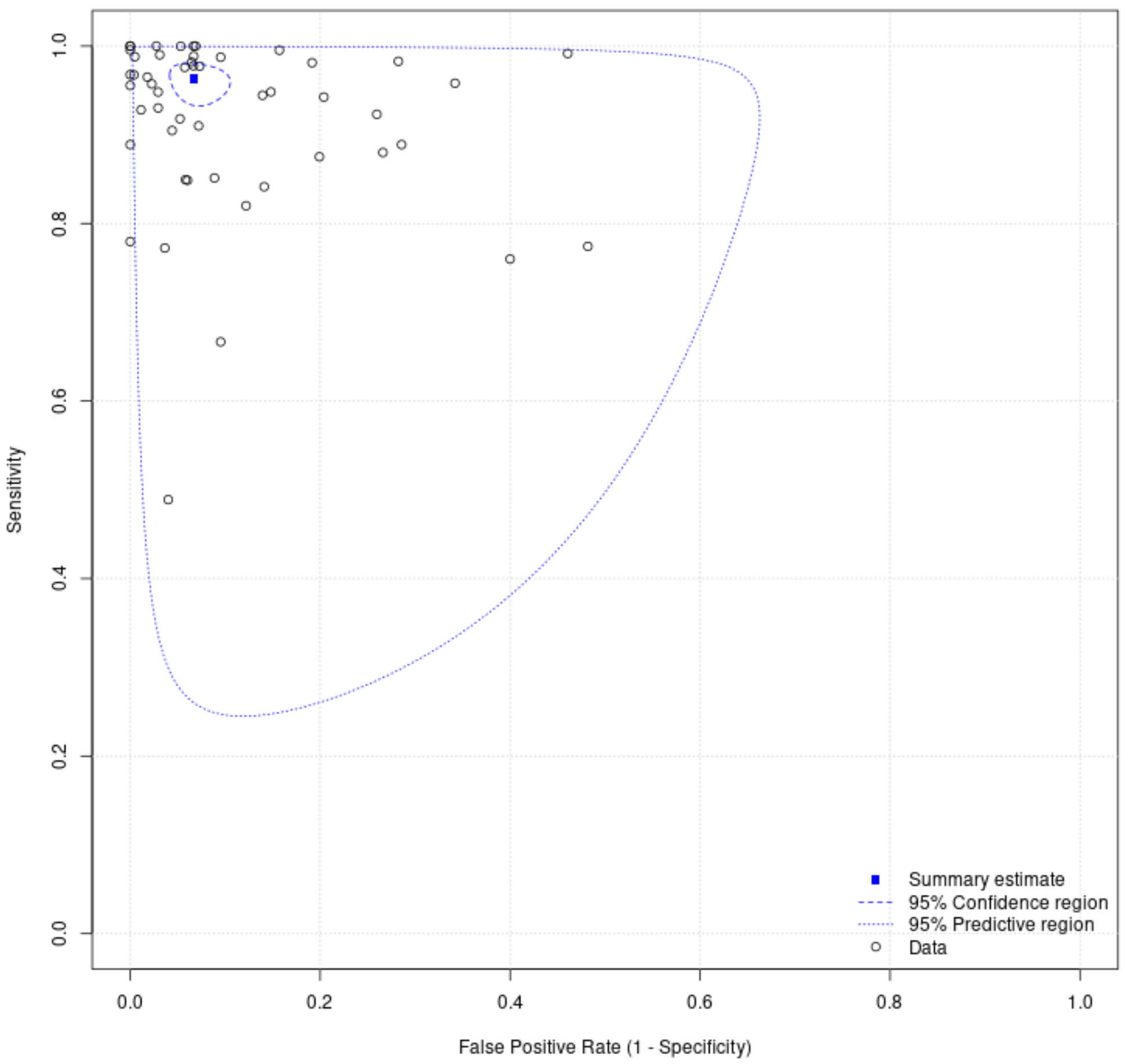

***Figure 5*** – Summary receiver operating characteristic plot of AI applied to whole slide images for all disease types generated from MetaDTA: Diagnostic Test Accuracy Meta-Analysis v2.01 Shiny App https://crsu.shinyapps.io/dta_ma/.[32,33]. 95% confidence intervals are shown around the summary estimate. The predictive region shows the area of 95% confidence in which the true sensitivity and specificity of future studies lies, whilst factoring the statistical heterogeneity of studies demonstrated in this review.

The largest subgroups of studies available for inclusion in the meta-analysis were studies of gastrointestinal pathology[36-48], breast pathology[35,49-55] and urological pathology[35,56-62] which are shown in ***Table 8***, representing over 60% of models included in the meta-analysis. Notably, studies of gastrointestinal pathology had a mean sensitivity of 93% and mean specificity of 94%. Similarly, studies of uropathology had mean sensitivities and specificities of 95% and 96% respectively. Studies of breast pathology had slightly lower performance at mean sensitivity of 83% and mean specificity of 88%. Results for all other disease types are also included in the meta-analysis.[63-82] Forest plots for these subgroups are shown in the supplementary materials. For studies that could not be included in the meta-analysis, an indication of best performance from other accuracy metrics provided is outlined in the supplementary materials.

Of models examined in the meta-analysis, the number of sources ranged from one to fourteen and overall the mean sensitivity and specificity improved with a larger number of data sources included in the study. For example, mean sensitivity and specificity for one data source was 89% and 88% respectively, whereas for three data sources this was 93% and 92% respectively. However, the majority of studies used one or two data sources only, meaning that studies with larger numbers of data sources were comparably underrepresented. Additionally, of these models, the mean sensitivity and specificity was higher in those validated on an external test set (95% and 92% respectively compared to those without external validation (91% and 87% respectively), although it must be acknowledged that frequently raw data was only available for internal validation performance. Similar performance was reported across studies that had a slide-level and patch / tile-level unit of analysis with a mean sensitivity of 95% and 91% respectively versus a mean specificity of 88% and 90% respectively. Further details of these findings can be found in the supplementary materials.

## DISCUSSION

AI has been extensively promoted as a useful tool that will transform medicine, with examples of innovation in clinical imaging, electronic health records (EHR), clinical decision making, genomics, wearables, drug development and robotics.[83-88] The potential of AI in digital pathology has been identified by many groups, with discoveries frequently emerging and attracting considerable interest.[9,89] Tools have not only been developed for diagnosis and prognostication, but also for predicting treatment response and genetic mutations from the H&E image alone.[8,9,11] Various models have now received regulatory approval for applications in pathology, with some examples being trialed in clinical settings.[62,90]

Despite the many interesting discoveries in pathology AI, translation to routine clinical use remains rare and there are many questions and challenges around the evidence quality, risk of bias and robustness of the medical AI tools in general.[22-24,91,92] This is the first systematic review and meta-analysis to address the diagnostic accuracy of AI models for detecting disease in digital pathology across all disease areas. It is a broad review of the performance of pathology AI, addresses the risk of bias in these studies, highlights the current gaps in evidence and also the deficiencies in reporting of research. Whilst the authors are not aware of a comparable study in pathology AI, Aggarwal et al. performed a similar review of deep learning in other (non-pathology) medical imaging types and found high diagnostic accuracy in ophthalmology imaging, respiratory imaging and breast imaging.[83] Whilst there are many exciting developments across medical imaging AI, ensuring that products are accurate and underpinned by robust evidence is essential for their future clinical utility and patient safety.

### Findings

This study sought to determine the diagnostic test accuracy of artificial intelligence solutions applied to whole slide images to diagnose disease. Overall, the meta-analysis showed that AI has a high sensitivity and specificity for diagnostic tasks across a variety of disease types in whole slide images (***Figure 4***). The performance of the models described in studies that were not included in the meta-analysis were also promising (see supplementary materials).

Subgroups of gastrointestinal pathology, breast pathology and urological pathology studies were examined in more detail, as these were the largest subsets of studies identified (see ***Table 8*** and supplementary materials). The gastrointestinal subgroup demonstrated high mean sensitivity and specificity and included AI models for colorectal cancer [36-38,40,42,48], gastric cancer[36,39,41,45-47,93] and gastritis[43]. The breast subgroup included only AI models for breast cancer applications, with Hameed et al. and Wang et al. demonstrating particularly high sensitivity (98%, 91% respectively) and specificity (93%, 96% respectively).[50,53] However, there was lower diagnostic accuracy in the breast group compared to some other specialties. This could be due to several factors, including challenges with tasks in breast cancer itself, an over-estimation of performance and bias in other areas and the differences in datasets and selection of data between subspecialty areas. Overall results were most favourable for the subgroup of urological studies with both high mean sensitivity and specificity (***Table 8***). This subgroup included models for renal cancer[56,60] and prostate cancer[35,57-59,61,62]. Whilst high diagnostic accuracy was seen in other subspecialties (***Table 8***), for example mean sensitivity and specificity in neuropathology (100%, 95% respectively) and soft tissue and bone pathology (98%, 94% respectively), there were very few studies in these subgroups and so the larger subgroups are likely more representative.

Of studies of other disease types included in the meta-analysis (***Figure 4***), AI models in liver cancer[82], lymphoma[81], melanoma[80], pancreatic cancer[79], brain cancer[75] lung cancer[65] and rhabdomyosarcoma[64] all demonstrated a high sensitivity and specificity. This emphasises the breadth of potential diagnostic tools for clinical applications with a high diagnostic accuracy in digital pathology.

Sensitivity and specificity were higher in studies with a greater number of included data sources, however few studies chose to include more than two sources of data. To develop AI models that can be applied in different institutions and populations, a diverse dataset is an important consideration for those conducting research into models intended for clinical use. A higher mean sensitivity and specificity for those models that included an external validation was identified, although many studies did not include this, or included most data for internal validation performance. Improved overall reporting of these values would allow a greater understanding of the performance of models at external validation. Performance was similar in the models included in the meta-analysis when a slide-level or patch / tile-level analysis was performed, although slide-level performance could be more useful when interpreting the clinical implications of a proposed model.

## Limitations

It must be acknowledged that there is uncertainty in the interpretation of the diagnostic accuracy of the AI models demonstrated in these studies. There was substantial heterogeneity in the study design, metrics used to demonstrate diagnostic accuracy, size of datasets, unit of analysis (e.g. slide, patch, pixel, specimen) and the level of detail given on the process and conduct of the studies. For instance, the total number of WSIs used in the studies for development and testing of AI models ranged from less than ten WSIs to tens of thousands of WSIs.[94,95] Of the 100 papers identified for inclusion in this review, 99% had at least one area at high or uncertain risk of bias, meaning any results should be interpreted with caution. Many studies had multiple areas at risk of bias and applicability concerns (***Figure 3***).

Whilst 100 papers were identified, only 48 studies were included in the meta-analysis due to deficient reporting. Whilst the meta-analysis provided a useful indication of diagnostic accuracy across disease areas, data for true positive, false positive, false negative and true negative was frequently missing and therefore made the assessment more challenging. To address this problem, missing data was requested from authors. Where a multiclass study output was provided, this was combined into a 2x2 confusion matrix to reflect disease detection / diagnosis, however this offers a more limited indication of diagnostic accuracy. AI specific reporting guidelines for diagnostic accuracy should help to improve this problem in future.[31]

Diagnostic accuracy in many of the described studies was high. There is likely a risk of publication bias in the studies examined, with poorer performing models not appearing in the literature. AI research is especially at risk of this, given it is currently a fast moving and competitive area. Many studies either used datasets that were not randomly selection or representative of the general patient population, or were unclear in their description of case selection, meaning studies were at risk of selection bias. The majority of studies used either one or two data sources only and therefore the training and test datasets may have been comparatively similar. All of these factors should be considered when interpreting performance.

## Conclusions

There are many promising applications for AI models in WSIs to assist the pathologist. This systematic review has outlined a high diagnostic accuracy for AI across multiple disease types. A larger body of evidence is available for gastrointestinal pathology, urological pathology and breast pathology. Many other disease areas are underrepresented and should be explored further in future. To improve the quality of future studies, reporting of sensitivity, specificity and raw data (true positives, false positives, false negatives, true negatives) for pathology AI models would help with transparency in comparing diagnostic performance between studies. Providing a clear outline of the breakdown of data and the data sources used in model development and testing would improve interpretation of results and transparency. Performing an external validation on data from an alternative source to that on which an AI model was trained, providing details on the process for case selection and using large, diverse datasets would help to reduce the risk of bias of these studies. Overall, better quality study design, transparency, reporting quality and addressing substantial areas of bias is needed to improve the evidence quality in pathology AI and to therefore harness the benefits of AI for patients and clinicians.

## COMPETING INTERESTS & ACKNOWLEDGEMENTS

The authors declare that there are no competing interests.

Dr McGenity, Dr Jennings, Dr Matthews and Prof. Treanor are funded by the National Pathology Imaging Co-operative (NPIC). NPIC (project no. 104687) is supported by a £50m investment from the Data to Early Diagnosis and Precision Medicine strand of the Government's Industrial Strategy Challenge Fund, managed and delivered by UK Research and Innovation (UKRI).

Dr Clarke is supported by the Medical Research Council (MR/S001530/1) and the Alan Turing Insititute.

Dr Cartlidge is supported by the National Institute for Health and Care Research (NIHR) Leeds Biomedical Research Centre. The views expressed are those of the authors and not necessarily those of the NHS, the NIHR or the Department of Health and Social Care.

Mr Freduah-Agyemang is supported by the EXSEL Scholarship Programme at the University of Leeds.

The funders had no role in the study design, data collection, data analysis or writing the manuscript.

We thank the authors who kindly provided additional data for the meta-analysis.

## DATA AVAILABILITY

All data generated or analysed during this study are included in this published article and its supplementary information files.

## AUTHOR CONTRIBUTIONS

CM, ELC, DT and DDS planned the study. CM conducted the searches. Abstracts were screened by CM and HFA. Full text articles were screened by CM and ELC. Data extraction was performed by CM, ELC, CJ, GM and CC. CM analysed the data and wrote the manuscript, which was revised by ELC, CJ, GM, CC, HFA, DDS and DT.

**Table 1.** Characteristics of breast pathology studies

| First author, year & reference | Location | Index test | Disease studied | Reference standard | Data sources | Training set details | Validation set details | Test set details | External validation | Unit of analysis |
|---|---|---|---|---|---|---|---|---|---|---|
| Cengzig (2022)[55] | Turkey | CNN | Breast cancer | Not stated | Not stated | 296,675 patches | | 101,706 patches | Unclear | Patch / Tile |
| Choudhary (2021)[54] | India, USA | CNN (VGG19, ResNet54, ResNet50) | Breast cancer | Pathologist annotations, slide diagnoses | IDC dataset | 194,266 patches | | 83,258 patches | No | Patch / Tile |
| Cruz-Roa (2018)[96] | Colombia, USA | FCN (HASHI) | Breast cancer | Pathologist annotations | Hospital of the University of Pennsylvania; University Hospitals Case Medical Centre / Case Western Reserve University; Cancer Institute of New Jersey; TCGA | 698 cases | 52 cases | 195 cases | Yes | Pixel |
| Cruz-Roa (2017)[97] | Colombia, USA | CNN (ConvNet) | Breast cancer | Pathologist annotations | University of Pennsylvania Hospital; University Hospitals Case Medical Centre / Case Western Reserve University; Cancer Institute of New Jersey; TCGA | 349 patients | 40 patients | 216 patients | Yes | Pixel |
| Hameed (2020)[53] | Spain, Columbia | CNN (ensemble of fine-tuned VGG16 & fine-tuned VGG19) | Breast cancer | Pathologist labels & annotations | Colsanitas Colombia University | 540 images/patches | 135 images/patches | 170 images/patches | No | Patch / Tile |
| Jin (2020)[52] | Canada | U-net CNN (ConcatNet) | Breast cancer | Labels | PatchCamelyon dataset; Open-source dataset from PMID 27563488; Warwick dataset | 262,144 patches + 538 images | 32,768 patches | 32,768 patches | No | Patch / Tile |
| Johny (2021)[98] | India | Custom deep CNN | Breast cancer | Pathologist patch labels | PatchCamelyon Dataset | 262,144 patches | | 65,536 patches | No | Patch / Tile |
| Kanavati (2021)[51] | Japan | CNN tile classifier (EfficientNetB1) + RNN tile aggregator | Breast cancer | Diagnostic review by pathologists | International University of Health and Welfare, Mita Hospital; Sapporo-Kosei General Hospital. | 1652 WSIs | 90 WSIs | 1930 WSIs | Yes | Slide |
| Khalil (2022)[99] | Taiwan | Modified FCN | Breast cancer | Pathologist annotations, IHC. | National Taiwan University Hospital dataset | 68 WSIs | | 26 WSIs | No | Slide |
| Lin (2019)[100] | Hong Kong, China, UK | Modified FCN | Breast cancer | Slide level labels, pathologist annotations | Camelyon dataset | 202 WSIs | 68 WSIs | 130 WSIs | No | Slide |
| Roy (2021)[101] | India, Germany | Multiple machine learning classifiers (CatBoost & others) | Breast cancer | Unclear | IDC Breast Histopathology Image Dataset | | Unclear | | No | Patch / Tile |
| Sadeghi (2019)[102] | Germany, Austria | CNN | Breast cancer | Pathologist supervised annotations, IHC | Camelyon17 dataset; Camelyon16 dataset | 400 WSIs | 100 WSIs | 20,000 patches | No | Patch / Tile |
| Steiner (2018)[103] | USA | CNN (LYNA - Inception framework) | Breast cancer | Pathologist review, IHC | Camelyon; Expired clinical archive blocks from 2 sources | 215 WSIs | 54 WSIs | 70 WSIs | Yes | Slide |
| Valkonen (2017)[104] | Finland | Random forest | Breast cancer | Pathologist WSI annotations | Camelyon16 dataset | 1,000,000 patches | 270 WSIs leave-one-out cross validation | | Yes | Patch / Tile |
| Wang Q (2021)[50] | China | SoMIL) + adaptive aggregator + RNN | Breast cancer | WSI labels, pixel level annotations of metastases | Camelyon16; MSK breast cancer metastases dataset | 289 WSIs | | 240 WSIs | Yes | Slide |
| Wu (2020)[49] | USA | ROI classifier + Tissue segmentation CNN + Diagnosis classifier SVM | Breast cancer | Pathologist pixel labels | Breast Cancer Surveillance Consortium–associated tumor registries in New Hampshire and Vermont | 58 ROIs | Cross validation 428 ROIs | | Unclear | Other (ROIs) |

**Table 2.** Characteristics of cardiothoracic pathology studies

| First author, year & reference | Location | Index test | Disease studied | Reference standard | Data sources | Training set details | Validation set details | Test set details | External validation | Unit of analysis |
|---|---|---|---|---|---|---|---|---|---|---|
| Chen (2021)[105] | Taiwan | CNN | Lung cancer | Pathologist diagnosis,slide level labels. | Taipei Medical University Hospital; Taipei Muncipal Wanfang Hospital; Taipei Medical University Shuang-Ho Hospital; TCGA. | 5045 WSIs | 561 WSIs | 2441 WSIs | Yes | Slide |
| Chen (2022)[106] | China | CNN (EfficientNetB5) | Lung cancer | Pathologist annotations | Hospital of Sun Yat-sen University; Shenzhen People's Hospital; Cancer Centre of Guangzhou Medical University | 813 cases train & validate | | 1101 cases | Yes | Slide |
| Coudray (2018)[107] | USA, Greece | CNN (Inception v3) | Lung cancer | Pathologist labels | TCGA, New York University | 1157 WSIs | 234 WSIs | 584 WSIs | Yes | Slide |
| Dehkharghanian (2021)[108] | Canada, USA | DNN (KimiaNet) | Lung cancer | WSI diagnostic label | TCGA; Grand River Hospital, Kitchener, Canada. | 575 WSIs | 79 WSIs | 81 WSIs | Yes | Patch / Tile |
| Kanavati (2020)[76] | Japan | CNN (EfficientNet-B3) | Lung cancer | Pathologist review & annotations | Kyushu Medical Centre; Mita Hospital; TCGA; TCIA | 3554 WSIs | 150 WSIs | 2170 WSIs | Yes | Slide |
| Wang X (2020)[65] | China, Hong Kong, UK | FCN + Random Forest classifier | Lung cancer | Pathologist annotations, WSI labels. | Sun Yat-sen University Cancer Centre (SUCC); TCGA | 1154 WSIs | | 285 WSIs | Yes | Slide |
| Wei (2019)[109] | USA | CNN (ResNet) | Lung cancer | Pathologist WSI labels | Dartmouth-Hitchcock Medical Centre (DHMC) | 245 WSIs | 34 WSIs | 143 WSIs | No | Slide |
| Yang (2021)[110] | China | CNN (EfficientNetB5; ResNet50) | Lung cancer | Pathologist diagnosis, IHC, medical records. | Sun Yat-sen University; Shenzhen People's Hospital; TCGA | 511 WSIs | 115 WSIs | 1067 WSIs | Yes | Patch / Tile |
| Zhao (2021)[63] | China | Combined (MR-EM-CNN + HMS + RNN + RMDL) | Lung cancer | Pathologist annotations, patch labels. | TCGA | 1481 WSIs | 321 WSIs | 323 WSIs | No | Slide |
| Zheng (2022)[111] | USA | CNN (GTP: Graph transformer + node representation connectivity information + feature generation & contrastive learning) | Lung cancer | Pathologist annotations, WSI level labels. | Clinical Proteomic Tumor Analysis Consortium (CPTAC), TCGA; the National Lung Screening Trial (NLST) | 2071 WSIs 5 fold cross validation | | 2082 WSIs | Yes | Slide |
| Uegami (2022)[112] | Japan | CNN (ResNet18) + K means clustering + pathologist clustering + transfer learning | Interstitial lung disease | Pathologist diagnosis | 1 institute (unclear) | 126 cases | 54 cases | 180 WSIs (51 cases) | No | Patch / Tile |

**Table 3.** Characteristics of dermatopathology studies

| First author, year & reference | Location | Index test | Disease studied | Reference standard | Data sources | Training set details | Validation set details | Test set details | External validation | Unit of analysis |
|---|---|---|---|---|---|---|---|---|---|---|
| Kimeswenger (2020)[113] | Austria, Switzerland | CNN + ANN (Feature constructor ImageNet CNN + classification ANN) | Basal cell carcinoma | Categorised by pathologist | Kepler University Hospital; Medical University of Vienna. | 688 WSIs | | 132 WSIs | No | Patch / Tile |
| Alheejawi (2021)[94] | Canada, India | CNN | Melanoma | MART-1 stained images | University of Alberta, Canada | 70 960x960 pixel images | 15 960x960 pixel images | 15 960x960 pixel images | No | Pixel |
| De Logu (2020)[80] | Italy | CNN (Inception ResNet v2) | Melanoma | Pathologist review | University of Florence; University Hospital of Siena; Institute of Biomolecular Chemistry, National research Council | 45 WSIs | 15 WSIs | 40 WSIs | No | Patch / Tile |
| Hekler (2019)[78] | Germany | CNN (ResNet50) | Melanoma | Image labels | Dr Dieter Krahl institute, Heidelberg | 595 cropped images | | 100 cropped images | No | Patch / Tile |
| Hohn (2021)[77] | Germany | CNN (ResNeXt50) | Melanoma | Pathologist diagnosis | Two laboratories unspecified | 232 WSIs | 67 WSIs | 132 WSIs | No | Slide |
| Li (2021)[114] | China | CNN (ResNet50) | Melanoma | Pathologist WSI annotations | Central South University Xiangya Hospital; TCGA | 491 WSIs | 105 WSIs | 105 WSIs | No | Slide |
| Wang L (2020)[66] | China | CNN for patch-level classification (VGG16) & random forest for WSI-level classification | Melanoma | Pathologist diagnosis, consensus, IHC, annotations. | Zhejiang University School of Medicine; Ninth People's Hospital of Shanghai | 105,415 patches | 1962 patches | 118,123 patches | Yes | Patch / Tile |
| del Amor (2021)[115] | Spain | CNN (VGG16, ResNet50, InceptionV3, MobileNetV2) | Spitzoid skin tumours | Pathologist annotations | CLARIFYv1 | 36 WSIs<br>5 fold cross validation of training set | | 15 WSIs | No | Unclear |

**Table 4.** Characteristics of hepatobiliary pathology studies

| First author, year & reference | Location | Index test | Disease studied | Reference standard | Data sources | Training set details | Validation set details | Test set details | External validation | Unit of analysis |
|---|---|---|---|---|---|---|---|---|---|---|
| Aatresh (2021)[82] | India | CNN (LiverNet) | Liver cancer | Pathologist annotations | Kasturba Medical College (KMC); TCGA | 5 fold cross-validation 5450 samples | | | No | Patch / Tile |
| Chen (2020)[116] | China | CNN (Inception V3) | Liver cancer | Labels | TCGA, Sir Run-Run Shaw Hospital | 278 WSIs | 56 WSIs | 258 WSIs | Yes | Patch / Tile |
| Kiani (2020)[117] | USA | CNN (Densenet) | Liver cancer | Pathologist diagnosis, consensus, IHC, special stains | TCGA; Stanford whole-slide image dataset | 20 WSIs | 50 WSIs | 106 WSIs | Yes | Slide |
| Yang (2022)[118] | Taiwan | Feature Aligned Multi-Scale Convolutional Network (FA-MSCN) | Liver cancer | Pathologist labels and ROIs | Unclear | 20 WSIs | | 26 WSIs | Unclear | Unclear |
| Schau (2020)[70] | USA, Thailand | CNNs (Inception v4) | Liver metastases | Pathologist labels, annotations | OHSU Knight BioLibrary | 200 WSIs | | 85 WSIs | No | Patch / Tile |
| Fu (2021)[79] | China | CNN (InceptionV3 patch-level classification), lightGBM model (WSI-level classification) & U-Net CNN (patch-level segmentation) | Pancreatic cancer | Pathologist annotations, labels | Peking Union Medical College Hospital (PUMCH); TCGA | 79,588 patches | 9952 patches | 9,948 patches + 52 WSIs | Yes | Slide |
| Naito (2021)[71] | Japan | CNN (EfficientNetB1) | Pancreatic cancer | Pathologist review, pathologist annotations | Kurume University | 372 WSIs | 40 WSIs | 120 WSIs | No | Slide |
| Song (2013)[68] | South Korea | Bayesian classifier; k-NN; SVM; ANN. | Pancreatic neoplasms | Unclear | Pathology department of Yeognam University | 240 patches | | 160 patches | No | Patch / Tile |

**Table 5.** Characteristics of gastrointestinal pathology studies

| First author, year & reference | Location | Index test | Disease studied | Reference standard | Data sources | Training set details | Validation set details | Test set details | External validation | Unit of analysis |
|---|---|---|---|---|---|---|---|---|---|---|
| Sali (2020)[119] | USA | CNN & Random forest; SVM; k-means; GMM | Barrett's Oesophagus | Pathologist consensus, pixel-wise annotations | Hunter Holmes McGuire Veterans Affairs Medical Center | 115 WSIs | 535 WSIs 10 fold cross validation | | No | Slide |
| Syed (2021)[120] | USA, Pakistan, Zambia, UK | CNN (ResNet50; ResNet50 multi-zoom; shallow CNN; ensemble). | Coeliac & Environmental Enteropathathy | Slide level diagnosis, IHC, patch labels. | Aga Khan University; University of Zambia & University Teaching Hospital Zambia; University of Virginia, USA | 231 WSIs | 115 WSIs | 115 WSIs | Unclear | Slide |
| Nasir-Moin (2021)[121] | USA | CNN (ResNet18) | Colorectal adenoma / polyps | Pathologist consensus | Dartmouth-Hitchcock Medical Centre (DHMC). Prior validation on 24 US institutions | 508 WSIs | | 100 WSIs + Previous validation 238 WSIs | Yes | Slide |
| Song (2020a)[44] | China | CNN (DeepLab v2 + ResNet34) | Colorectal adenoma / polyps | Pathologist labels | Chinese People's Liberation Army General Hospital (PLAGH); China-Japan Friendship Hospital (CJFH); Cancer Hospital, Chinese Academy of Medical Science (CH). | 177 WSIs | 40 WSIs | 362 WSIs | Yes | Slide |
| Wei (2020)[122] | USA | CNN (ResNet) | Colorectal adenoma / polyps | Pathologist annotations | Dartmouth-Hitchcock Medical Centre (DHMC); External set multiple institutions | 326 WSIs | 25 WSIs | 395 WSIs | Yes | Slide |
| Feng (2021)[123] | China, USA, South Korea | CNN (ensemble of 8 networksmodified U-Net + VGG-16 or VGG-19) | Colorectal cancer | Pixel annotations, pathologist labels | DigestPath 2019 Challenge (task 2) | 750 WSIs | | 250 WSIs | No | Unclear |
| Haryanto (2021)[124] | Indonesia | Conditional Sliding Window (CSW) algorithm used to generate images for CNN 7-5-7 | Colorectal cancer | Pathologist labels & annotations | Warwick dataset; University of Indonesia | | Unclear | | Unclear | Unclear |
| Sabol (2020)[125] | Slovakia, Japan | CNN + X-CFCMC | Colorectal cancer | Annotations | Publicly available dataset from Kather et al. | | 10 fold cross validation 5000 tiles | | No | Patch / Tile |
| Schrammen (2022)[126] | Germany, Netherlands, UK | Single neural network (SLAM - based on ShuffleNet) | Colorectal cancer | Patient/slide level labels | DACHS study, YCR-BCIP | 2448 cases | | 889 cases | Yes | Slide |
| Tsuneki (2021)[42] | Japan | CNN (EfficientNetB1) | Colorectal cancer | Pathologist diagnosis & annotations | Wajiro, Shinmizumaki, Shinkomonji, & Shinyukuhashi hospitals, Fukuoka; Mita Hospital, Tokyo | 680 WSIs | 68 WSIs | 1799 WSIs | Yes | Slide |
| Wang KS (2017)[48] | China, USA | CNN (Inception V3) | Colorectal cancer | Pathologist consensus & labels | 14 hospitals / sources | 559 WSIs | 283 WSIs | At least 13,838 WSIs | Yes | Patch / Tile |
| Wang C (2017)[40] | China | CNN (bilinear) | Colorectal cancer | Annotations | University Medical Center Mannheim, Heidelberg | 5 fold cross validation on 1000 patches | | | No | Patch / Tile |
| Xu (2021)[38] | China | Dual resolution deep learning network with self-attention mechanism (DRSANet) | Colorectal cancer | Pathologist annotations, Patch labels, Pathologist pixel annotations. | TCGA; Affiliated Cancer Hospital and Institute of Guangzhou Medical University (ACHIGMU) | 100,000 patches | 40,000 patches | 80,000 patches | Yes | Patch / Tile |
| Zhou (2021)[127] | China, Singapore | CNN (ResNet) + Random Forest | Colorectal cancer | Pathologist slide labels, reports, annotations & consensus | TGCA; Hospital of Zhejiang University; Hospital of Soochow University; Nanjing First Hospital | 950 WSIs | | 446 WSIs | Yes | Slide |
| Ashraf (2022)[47] | South Korea | CNN (DenseNet-201) | Gastric cancer | Pathologist review & annotations | Seegene Medical Foundation in South Korea; Camelyon | Primary model: 723 WSIs; LN model: 262,11 patches | Primary model: 91 WSIs; LN model: 32,768 patches | Primary model: 91 WSIs; LN model: 32,768 patches | No | Patch |
| Cho (2019)[46] | South Korea | CNN (AlexNet; ResNet50; Inception-v3) | Gastric cancer | Labels | TCGA-STAD; SSMH Seoul St. Mary's Hospital dataset | | 10 fold cross validation | | Yes | Slide |
| Ma (2020)[128] | China | CNN (modified InceptionV3) + random forest classifier | Gastric cancer | Pathologist annotations | Ruijin Hospital | 534 WSIs | 76 WSIs | 153 WSIs | No | Slide |
| Rasmussen (2020)[45] | Canada | CNN (DenseNet169) | Gastric cancer | Pathologist annotations | Queen Elizabeth II Health Sciences Centre & Dalhousie University; Sunnybrook Health Science Centre, University of Toronto | 14,266 patches | 1585 patches | 1785 patches | Yes | Patch / Tile |
| Song (2020b)[93] | China, USA | CNN (Multiple models); random forest | Gastric cancer | Pathologist pixel level annotations | PLAGH dataset; Multicentre dataset (PUMCH, CHCAMS & Pekin Union Medical College) | 2860 WSIs | 300 WSIs | 4993 WSIs | Yes | Slide |
| Tung (2022)[41] | Taiwan | CNN (YOLOv4) | Gastric cancer | Pathologist annotations | Taiwan Cancer Registry Database | 2200 image tiles | | 550 image tiles | No | Patch / Tile |
| Wang S (2019)[39] | China | Recalibrated multi-instance deep learning method (RMDL) | Gastric cancer | Pathologist pixel annotations | Sun Yat-sen University | 408 WSIs | | 200 WSIs | No | Slide |
| Ba (2021)[129] | China | CNN (ResNet50) | Gastritis | Pathologist review & pixel annotations | Chinese People's Liberation Army General Hospital | 1008 WSIs | 100 WSIs | 142 WSIs | No | Slide |
| Steinbuss (2020)[43] | Germany | CNN (Xception) | Gastritis | Diagnoses – modified Sydney Classification, pathologist annotations | Institute of Pathology, University Clinic Heidelberg | 825 patches | 196 patches | 209 patches | No | Patch / Tile |
| Iizuka (2020)[36] | Japan | CNN (InceptionV3 + max-pooling or RNN aggregator) | Multiple (Colorectal cancer & Gastric tumours) | Pathologist annotations | Hiroshima University Hospital dataset; Haradoi Hospital dataset; TCGA dataset | Stomach: 3,628 WSIs; Colon: 3,536 WSIs | | Stomach: 1,475 WSIs; Colon: 1,574 WSIs | Yes | Slide |

**Table 6.** Characteristics of urological pathology studies

| **First author, year & reference** | **Location** | **Index test** | **Disease studied** | **Reference standard** | **Data sources** | **Training set details** | **Validation set details** | **Test set details** | **External validation** | **Unit of analysis** |
|---|---|---|---|---|---|---|---|---|---|---|
| da Silva (2021)[62] | Brazil, USA | CNN (Paige Prostate 1.0) | Prostate cancer | Pathologist consensus, IHC | Instituto Mario Penna, Brazil | Prior study: trained on 2000 WSIs | | 661 WSIs (579 part specimens) | Yes | Other (part specimen level) |
| Duran-Lopez (2021)[130] | Spain | CNN (PROMETEO) + Wide and deep neural network | Prostate cancer | Pathologist pixel annotations | Pathological Anatomy Unit of Virgen de Valme Hospital, Spain | 5 fold cross validation | | 332 WSIs | No | Slide |
| Esteban (2019)[61] | Spain | Optical density granulometry-based descriptor + Gaussian processes | Prostate cancer | Pathologist pixel annotations | SICAPv1 database; Prostate cancer database by Gertych et al. | 60 WSIs 5 fold cross validation | | 19 WSIs + 593 patches | Yes | Patch / Tile |
| Han (2020a)[131] | Canada | Multiple ML approaches (Transfer learning with TCMs & others) | Prostate cancer | Pathologist annotations & supervision | Western University | 286 WSIs cross validation for train / test (leave one out) | | 13 WSIs | No | Patch / Tile |
| Han (2020b)[59] | Canada | Traditional ML and 14 texture features extracted from TCMs; Transfer learning with pretrained AlexNet fine-tuned by TCM ROIs; Transfer learning with pretrained AlexNet fine-tuned with raw image ROIs | Prostate cancer | Pathologist annotations & supervision | Western University | 286 WSIs cross validation for train / test (leave one out) | | 13 WSIs | No | Patch / Tile |
| Huang (2021)[132] | USA | CNN (U-Net gland segmenter) + CNN feature extractor & classifier | Prostate cancer | Pathologist review, patch annotations using ISUP criteria. | University of Wisconsin Health System | 838 WSIs | | 162 WSIs | No | Other (patch-pixel level) |
| Swiderska-Chadaj (2020)[58] | Netherlands, Sweden | CNN (U-Net, DenseNetFCN, EfficientNet) | Prostate cancer | Slide level labels, pathologist annotations | The Penn State Health Department of Pathology; PAMM Laboratorium voor Pathologie; Radboud University Medical Center. | 264 WSIs | 60 WSIs | 297 WSIs | Yes | Slide |
| Tsuneki (2022)[57] | Japan | Transfer learning (TL-colon poorly ADC-2 (20x,512)); CNN (EfficientNetB1 20x, 512); CNN (EfficientNetB1 (10x,224) | Prostate cancer | Pathologist diagnosis & consensus | Wajiro, Shinmizumaki, Shinkomonji, and Shinyukuhashi hospitals, Fukuoka; TGCA | 1122 WSIs | 60 WSIs | 2512 WSIs | Yes | Slide |
| Abdeltawab (2021)[133] | USA, UAE | CNN (pyramidal) | Renal cancer | Pathologist review & annotations | Indiana University, USA | 38 WSIs | 6 WSIs | 20 WSIs | No | Pixel |
| Fenstermaker (2020)[60] | USA | CNN | Renal cancer | Pathology report | TCGA | 15,168 patches train / validate | | 4,286 patches | No | Patch / Tile |
| Tabibu (2019)[134] | India | CNNs (ResNet18 & 34) + SVM (DAG-SVM) | Renal cancer | Clinical information including pathology reports | TCGA | 1474 WSIs | 317 WSIs | 314 WSIs | Yes | Slide |
| Zhu (2021)[56] | USA | CNN (ResNet-18) + Decision Tree | Renal cancer | Pathologist annotations | Dartmouth-Hitchcock Medical Centre (DHMC); TCGA | 385 WSIs | 23 WSIs | 1074 WSIs | Yes | Slide |

| Table 7. Characteristics of other pathology / multiple pathology studies | | | | | | | | | | |
|---|---|---|---|---|---|---|---|---|---|---|
| **First author, year & reference** | **Location** | **Index test** | **Disease studied** | **Reference standard** | **Data sources** | **Training set details** | **Validation set details** | **Test set details** | **External validation** | **Unit of analysis** |
| BenTaieb (2017)[135] | Canada | K means + LSVM | Ovarian cancer | Pathologist consensus | Not stated | 68 WSIs | | 65 WSIs | No | Slide |
| Shin (2020)[69] | South Korea | CNN (Inception V3) | Ovarian cancer | Pathologist diagnosis | TCGA; Ajou University Medical Centre | 7245 patches | | 3051 patches | Yes | Patch / Tile |
| Sun (2020)[67] | China | CNN (HIENet) | Endometrial cancer | Pathologist consensus, patch labels | 2 datasets from Hospital of Zhenghou University | 10 fold cross validation on 3300 patches | | 200 patches | No | Patch / Tile |
| Yu (2020)[136] | USA | CNN (VGGNet, GoogLeNet; AlexNet) | Ovarian cancer | Pathology reports and pathologist review | TCGA | 1100 WSIs | | 275 WSIs | No | Slide |
| Achi (2019)[81] | USA | CNN | Lymphoma | Labels | Virtual pathology at University of Leeds, Virtual Slide Box University of Iowa | 1856 patches | 464 patches | 240 patches | No | Patch / Tile |
| Miyoshi (2020)[73] | Japan, USA | deep neural network classifier with averaging method | Lymphoma | Pathologist annotations, IHC | Kurume University | Unclear | Unclear | 100 patches | No | Patch / Tile |
| Mohlman (2020)[72] | USA | deep densely connected CNN | Lymphoma | Unclear - likely slide diagnosis | University of Utah dataset, Mayo Clinic Rochester dataset | 8796 patches | | 2037 patches | No | Patch / Tile |
| Syrykh (2020)[137] | France | CNNs ("Several Deep CNNs" + Bayesian Neural Network) | Lymphoma | Slide diagnosis, IHC, patch labels | Toulouse University Cancer Institute, France; Dijon University Hospital, France. | 221 WSIs | 111 WSIs | 159 WSIs | No | Slide |
| Yu (2019)[138] | USA | CNN (VGGNet & others) | Lymphoma | Pathologist consensus, IHC | TCGA & International Cancer Genome Consortium (ICGC) | 707 patients | | 302 patients | Yes | Patch / Tile |
| Yu (2021)[139] | Taiwan | HTC-RCNN (ResNet50). Decision-tree-based machine learning algorithm, XGBoost | Lymphoma | Pathologist diagnosis with WHO criteria, pathologist annotations | 17 hospitals in Taiwan (names not specified) | Detect: 27 ROIs. Classify 3 fold validation from 40 WSIs | Detect: 2 ROIs. Classify: 3 fold validation from 40 WSIs | Detect: 3 ROIs. Classify: 3 fold validation from 40 WSIs | Unclear | Slide |
| Li (2020)[74] | China, USA | CNN (Inception V3) | Thyroid neoplasms | Pathologist review | Peking Union Medical College Hospital | 279 WSIs | 70 WSIs | 259 WSIs | No | Slide |
| Xu (2017)[140] | China | CNN (AlexNet) + SVM classifier | Multiple (Brain tumours, colorectal cancer) | MICCAI brain: Labels Colorectal: Pathologist review & image crops | MICCAI 2014 Brain Tumor Digital Pathology Challenge & colon cancer dataset | Brain:80 images ; Colon: 359 cropped images | | Brain: 61 images; Colon: 358 cropped images | No | Patch / Tile |
| DiPalma (2021)[141] | USA | CNN (Resnet architecture but trained from scratch) | Multiple (Coeliac, lung cancer, renal cancer) | RCC & Coeliac: Pathologist diagnosis, Lung: pathologist annotations | TCGA, Darmouth-Hitchcock Medical Centre | Coeliac: 5908 tissue pieces; Lung: 239 WSIs, 2083 tissue pieces; Renal: 617 WSIs, 834 tissue pieces. | Coeliac: 1167 tissue pieces; | Coeliac: 25,284 tissue pieces; Lung: 34 WSIs, 305 tissue pieces; Renal: 265 WSIs, 364 tissue pieces. | No | Slide |
| Litjens (2016)[35] | Netherlands | CNN | Multiple (Prostate cancer; Breast cancer) | Pathologist annotations / supervision, pathology reports. | 3 datasets from Radboud University Medical Centre | Prostate: 100 WSIs; Breast: 98 WSIs. | Prostate: 50 WSIs; Breast: 33 WSI. | Prostate: 75 WSIs; Breast: 42 WSIs + Consecutive set: 98 WSIs | No | Slide |
| Menon (2022)[142] | India | FCN (ResNet18) | Multiple cancer types | Slide labels | TCGA | 6855 WSIs | 1958 WSIs | 979 WSIs | No | Patch / Tile |
| Noorbakhsh (2020)[95] | USA | CNN (InceptionV3) | Multiple cancer types | Pathologist annotations | TCGA, CPTAC. | 19,470 WSIs | | 10,460 WSIs | Yes | Slide |
| Yan (2022)[37] | China | Contrastive clustering algorithm to train CNN encoder + recursive cluster refinement method | Multiple (colorectal cancer / polyps, breast cancer) | NCT-CRC Patch classification, CAMELYON16 annotations. In-house: pathologist diagnosis | NCT-CRC dataset; Camelyon16 dataset; In-house colon polyp WSI dataset | NCT-CRC 80,000 patches; Camelyon16 80,000 patches; | NCT-CRC 10,000 patches; Camelyon16 10,000 patches. | NCT-CRC + In house polyp dataset: 10,000 patches + 20 patients; CAMELYON16 10,000 patches | Yes | Patch / Tile |
| Li (2018)[75] | China | CNN (GoogleLeNet) | Brain cancer | Diagnosed WSIs | Huashan Hospital, Fudan University | 67 WSIs | | 139 WSIs | No | Patch / Tile |
| Schilling (2018)[143] | Germany | Voting ensemble classifier (logistic regression, SVM, decision tree & random forest) | Hirschsprung's disease | Pathologist diagnosis against criteria, IHC | Institute of Pathology, Friedrich-Alexander-University Erlangen Nurnberg, Germany | 172 WSIs | 58 WSIs | 77 WSIs | No | Unclear |
| Mishra (2017)[144] | USA | CNN (LeNet & AlexNet) | Osteosarcoma | Manual annotations by senior pathologists. | Unclear | 38,400 patches | 12,800 patches | 12,800 patches | No | Patch / Tile |
| Zhang (2022)[64] | USA | CNN (Inception V3) | Rhabdomyosarcoma | WSIs reviewed and classified by pathologist | Children's oncology group biobanking study | 56 WSIs | 12 WSIs | 204 WSIs | Unclear | Patch / Tile |

**Table 8.** Mean performance across studies by pathological subspecialty

| Pathological subspecialty | No. AI models | Mean sensitivity | Mean specificity |
|---|---|---|---|
| Gastrointestinal pathology | 14 | 93% | 94% |
| Breast pathology | 8 | 83% | 88% |
| Uropathology | 8 | 95% | 96% |
| Hepatobiliary pathology | 5 | 90% | 87% |
| Dermatopathology | 4 | 89% | 81% |
| Cardiothoracic pathology | 3 | 98% | 76% |
| Haematopathology | 3 | 95% | 86% |
| Gynaecological pathology | 2 | 87% | 83% |
| Soft tissue & bone pathology | 1 | 98% | 94% |
| Head & neck pathology | 1 | 98% | 72% |
| Neuropathology | 1 | 100% | 95% |

## Supplementary materials

### Contents

## S1 – Search strategy of three databases (PubMed, EMBASE & CENTRAL)

Pubmed
Limits (Humans, English)

| | | |
|---|---|---|
| 1 | digital pathol*.ti,ab. | 961 |
| 2 | whole slide image.ti,ab. | 176 |
| 3 | histopathol*.ti,ab. | 132,723 |
| 4 | artificial intelligence.ti,ab. | 12,011 |
| 5 | deep learning.ti,ab. | 14,760 |
| 6 | machine learning.ti,ab. | 31,756 |
| 7 | neural network.ti,ab. | 21,974 |
| 8 | computer vision.ti,ab. | 2,141 |
| 9 | support vector machine.ti,ab. | 8,911 |
| 10 | #1 OR #2 OR #3 | 133,553 |
| 11 | #4 OR #5 OR #6 OR #7 OR #8 OR #9 | 70,047 |
| 12 | #10 AND #11 | 1,279 |

Embase Classic+Embase

| | | |
|---|---|---|
| 1 | digital pathol*.ti,ab. | 1952 |
| 2 | whole slide image.ti,ab. | 408 |
| 3 | histopathol*.ti,ab. | 370,508 |
| 4 | artificial intelligence.ti,ab. | 23,908 |
| 5 | deep learning.ti,ab. | 31,614 |
| 6 | machine learning.ti,ab. | 67,418 |
| 7 | neural network.ti,ab. | 59,436 |
| 8 | computer vision.ti,ab. | 5,963 |
| 9 | support vector machine.ti,ab. | 19,875 |
| 10 | 1 or 2 or 3 | 372,380 |
| 11 | 4 or 5 or 6 or 7 or 8 or 9 | 166,702 |
| 12 | 10 and 11 | 2,628 |
| 13 | limit 12 to (human and english language and (embase or medline)) | 1,537 |

CENTRAL

| ID | Search Hits | |
|---|---|---|
| #1 | "digital pathol*" | 0 |
| #2 | "whole slide image" | 14 |
| #3 | histopathol* | 10,595 |
| #4 | "artificial intelligence" | 1,141 |
| #5 | "deep learning" | 729 |
| #6 | "machine learning" | 1,904 |
| #7 | "neural network" | 1,148 |
| #8 | "computer vision" | 116 |
| #9 | "support vector machine" | 376 |
| #10 | #1 OR #2 OR #3 | 10,603 |
| #11 | #4 OR #5 OR #6 OR #7 OR #8 OR #9 | 4,135 |
| #12 | #10 AND #11 | 180 |
| #13 | #12 in Trials | 160 |

## S2 – Screening tools for inclusion of articles

*S2a: Screening tool for abstracts*

| | | |
|---|---|---|
| 1. | Is this article an original research paper?<br>(i.e. not a review, conference abstract, commentary etc.) | No = Reject<br>Yes = Next question |
| 2. | Is education the primary focus of the article? | Yes = Reject<br>No = Next question |
| 3. | Is this article examining whole slide imaging?<br>(i.e. not other imaging modalities e.g. other pathology imaging technologies, radiological imaging, endoscopy etc.) | No = Reject<br>Yes = Next question |
| 4. | Is this article examining a surgical pathology / histopathology problem(s)? (i.e. not cytology, autopsy, toxicology, forensics, descriptions of new systems or collaborations) | No = Reject<br>Yes = Next question |
| 5. | Is this article examining artificial intelligence for whole slide imaging?<br>(i.e. not manual annotation etc.) | No = Reject<br>Yes = Next question |
| 6. | Is this study examining diagnosis of a disease?<br>(I.e. not determining only prognosis, treatment response, molecular status etc or focused on a purely quality / technical issue for WSI)) | No = Reject<br>Yes = Next question |
| 7. | Is this study measuring diagnostic accuracy?<br>(i.e. referring to accuracy or including accuracy statistics) | No = Reject<br>Yes = Next question |
| 8. | Is this a study of humans?<br>(i.e. not an animal based study) | No = Reject<br>Yes = Next question |
| 9. | Is this study written in English? | No = Reject<br>Yes = Accept |

*S2b: Screening tool for full text articles*

| | | |
|---|---|---|
| 1. | Is this article an original research paper? | No = Reject<br>Yes = Next question |
| 2. | Is education the primary focus of the article? | Yes = Reject<br>No = Next question |
| 3. | Is this article examining whole slide imaging?<br>(not other modalities and not combined with other modalities in the analysis) | No = Reject<br>Yes = Next question |
| 4. | Is this article examining a surgical pathology / histopathology problem(s)? | No = Reject<br>Yes = Next question |
| 5. | Is this article examining artificial intelligence for whole slide imaging? | No = Reject<br>Yes = Next question |
| 6. | Is this study examining diagnosis of a disease?<br>(detection of disease or classification of disease subtypes only) | No = Reject<br>Yes = Next question |
| 7. | Is this study measuring diagnostic accuracy? | No = Reject<br>Yes = Next question |
| 8. | Is this a study of humans? | No = Reject<br>Yes = Next question |
| 9. | Is this study written in English? | No = Reject<br>Yes = Accept |
| 10. | Does the ground truth use or imply use of human pathologist using H&E or IHC? | No = Reject<br>Yes = Accept |
| 11. | Does the article describe a grand challenge exercise with models from multiple authors?<br>(Rather than diagnostic accuracy study from one group) | Yes = Reject<br>No = Accept |

# S3 – QUADAS-2 tool tailored for this review

## Appendix – Adapted QUADAS2 tool

### Domain 1: Patient Selection

***Risk of Bias (describe methods of patient selection)***

Signaling questions

- Was a consecutive or random sample of cases used in the test set(s)? Yes/No/Unclear
- Did the study avoid inappropriate exclusions? (I.e. excluding all artefacts or excluding cases that were difficult to diagnose) Yes/No/Unclear

***QUESTION 1 - Could the selection of patients have introduced bias? RISK: LOW/HIGH/UNCLEAR***

Low risk (1) if all the answers to signalling questions were 'yes'
High risk (2) if any of the answers to signalling questions were 'no'
Unclear (3) if answer to signalling questions was 'unclear'

***Concerns regarding applicability (describe included patients)***

***QUESTION 2 – Is there a concern that the included patients do not match the review question? CONCERN: LOW/HIGH/UNCLEAR***

Low risk (1) if cases were selected from a given condition, without excluding subgroups
High risk (2) if subgroups of cases with a given condition were excluded, not reflecting the full case mix
Unclear (3) if it is not clear how cases were selected

### Domain 2: Index Test(s)

***Risk of bias (describe the index test and how it was conducted and interpreted)***

Signaling questions

- Were the reported performance results from test data that was independent of the training data? Yes/No/Unclear
- Was the index test tested on an external independent test set? Yes/No/Unclear
- Was the same image analysis performed on all the cases? Yes/No/Unclear
- Were all test cases used in the analysis? Yes/No/Unclear

***QUESTION 3 – Could the conduct or interpretation of the index test have introduced bias? RISK: LOW/HIGH/UNCLEAR***

Low risk (1) if all the answers to signalling questions were 'yes'
High risk (2) if any of the answers to signalling questions were 'no'
Unclear (3) if answer to signalling questions was 'unclear'

***Concerns regarding applicability***

***QUESTION 4 – Is there a concern that the index test, its conduct, or interpretation differ from the review question? CONCERN: LOW/HIGH/UNCLEAR***

Low risk (1) if there is no concern that the index test, its conduct or interpretation differ from the review question
High risk (2) if there is concern of either the index test, its conduct or interpretation differing from the review question
Unclear (3) if it is not clear if the index test, its conduct or interpretation differ from the review question.

### Domain 3: Reference Standard

***Risk of bias (describe the reference standard and how it was conducted and interpreted)***

Signaling questions

- Is the reference standard likely to correctly classify the target condition? Yes/No/Unclear
- Were the reference standard results interpreted without knowledge of the results of the index test? Yes/No/Unclear

***QUESTION 5 – Could the reference standard, its conduct, or its interpretation have introduced bias? RISK: LOW/HIGH/UNCLEAR***

Low risk (1) if the answers to both signalling questions were 'yes'
High risk (2) if the answers to either signalling questions were 'no'
Unclear (3) if answer to either signalling questions was 'unclear'

***Concerns regarding applicability***

***QUESTION 6 – Is there concern that the target condition as defined by the reference standard does not match the review question? CONCERN: LOW/HIGH/UNCLEAR***

Low risk (1) if the criteria for diagnosis was clearly defined and the target condition diagnosed by a pathologist.
High risk (2) if the criteria for diagnosis was not clearly defined or if the target condition was not diagnosed by a pathologist.
Unclear (3) if the criteria for diagnosis of a given condition was unclear or if it is not clear who diagnosed the target condition.

### Domain 4: Flow and Timing

***Risk of bias (describe the index test and how it was conducted and interpreted)***

Signaling questions

1. Was the time interval between diagnosis of the reference standard and the scanning of the glass slides for whole slide images <10 years? Yes/No/Unclear

***QUESTION 7 – Could the case flow have have introduced bias? RISK: LOW/HIGH/UNCLEAR***

Low risk (1) if answer to signalling question was 'yes'
High risk (2) if answer to signalling question was 'no'
Unclear (3) if answer to signalling question was 'unclear'

## S4 – Individual paper scores for QUADAS-2 assessment

| First author | Publication year | Risk of Bias: Patient selection | Risk of Bias: Index test | Risk of Bias: Reference standard | Risk of Bias: Flow and timing | Concerns of Applicability: Patient selection | Concerns of Applicability: Index test | Concerns of Applicability: Reference standard |
|---|---|---|---|---|---|---|---|---|
| Aatresh[82] | 2021 | 3 | 2 | 3 | 3 | 3 | 1 | 3 |
| Abdeltawab[133] | 2021 | 1 | 2 | 1 | 3 | 3 | 1 | 1 |
| Achi[81] | 2019 | 2 | 2 | 3 | 3 | 3 | 3 | 3 |
| Alheejawi[94] | 2021 | 3 | 2 | 3 | 1 | 3 | 3 | 3 |
| Ashraf[47] | 2022 | 2 | 2 | 1 | 3 | 3 | 1 | 1 |
| Ba[129] | | 3 | 2 | 1 | 3 | 3 | 1 | 1 |
| BenTaieb[135] | 2017 | 3 | 1 | 1 | 3 | 3 | 1 | 1 |
| Cengzig[55] | 2022 | 3 | 3 | 3 | 3 | 3 | 3 | 3 |
| Chen[105] | 2021 | 3 | 1 | 1 | 1 | 3 | 1 | 1 |
| Chen[116] | 2020 | 2 | 1 | 1 | 3 | 3 | 1 | 3 |
| Chen[106] | 2022 | 3 | 2 | 1 | 3 | 3 | 3 | 1 |
| Cho[46] | 2019 | 2 | 3 | 1 | 3 | 3 | 3 | 3 |
| Choudhary[54] | 2021 | 2 | 2 | 1 | 3 | 3 | 3 | 3 |
| Coudray[107] | 2018 | 2 | 2 | 1 | 3 | 3 | 1 | 1 |
| Cruz-Roa[96] | 2018 | 3 | 1 | 1 | 3 | 3 | 1 | 3 |
| Cruz-Roa[97] | 2017 | 3 | 1 | 1 | 3 | 3 | 1 | 3 |
| da Silva[62] | 2021 | 1 | 2 | 1 | 1 | 1 | 1 | 1 |
| De Logu[80] | 2020 | 3 | 2 | 1 | 3 | 3 | 3 | 1 |
| Dehkharghanian[108] | 2021 | 2 | 1 | 3 | 3 | 3 | 1 | 1 |
| del Amor[115] | 2021 | 3 | 2 | 1 | 3 | 3 | 1 | 1 |
| DiPalma[141] | 2021 | 3 | 2 | 3 | 3 | 3 | 1 | 2 |
| Duran-Lopez[130] | 2021 | 3 | 2 | 3 | 3 | 3 | 3 | 3 |
| Esteban[61] | 2019 | 2 | 1 | 1 | 3 | 3 | 1 | 3 |
| Feng[123] | 2021 | 2 | 2 | 3 | 3 | 3 | 1 | 2 |
| Fenstermaker[60] | 2020 | 2 | 2 | 3 | 3 | 3 | 3 | 1 |
| Fu | 2021 | 2 | 1 | 3 | 3 | 3 | 1 | 3 |
| Hameed[53] | 2020 | 3 | 2 | 3 | 3 | 3 | 1 | 2 |
| Han[131] | 2020a | 3 | 2 | 1 | 3 | 3 | 1 | 1 |
| Han[59] | 2020b | 3 | 2 | 1 | 3 | 3 | 1 | 1 |
| Haryanto[124] | 2021 | 2 | 2 | 3 | 3 | 3 | 2 | 2 |
| Hekler[78] | 2019 | 1 | 2 | 1 | 1 | 1 | 1 | 1 |
| Hohn[77] | 2021 | 3 | 2 | 1 | 3 | 3 | 3 | 1 |
| Huang[132] | 2021 | 1 | 2 | 1 | 2 | 1 | 1 | 1 |
| Iizuka[36] | 2020 | 3 | 1 | 1 | 3 | 3 | 1 | 1 |
| Jin[52] | 2020 | 2 | 2 | 3 | 3 | 3 | 1 | 2 |
| Johny[98] | 2021 | 2 | 2 | 3 | 3 | 3 | 1 | 2 |
| Kanavati[76] | 2020 | 3 | 1 | 1 | 3 | 3 | 1 | 1 |
| Kanavati[51] | 2021 | 3 | 1 | 1 | 3 | 3 | 1 | 1 |
| Khalil[99] | 2022 | 3 | 2 | 1 | 3 | 3 | 1 | 3 |
| Kiani[117] | 2020 | 1 | 1 | 1 | 1 | 1 | 1 | 2 |
| Kimeswenger[113] | 2020 | 3 | 2 | 1 | 3 | 3 | 1 | 1 |
| Li[114] | 2021 | 3 | 2 | 1 | 3 | 3 | 1 | 3 |
| Li[75] | 2018 | 3 | 2 | 3 | 3 | 3 | 1 | 2 |
| Li[74] | 2020 | 3 | 2 | 1 | 1 | 3 | 1 | 1 |
| Lin[100] | 2019 | 2 | 2 | 3 | 3 | 3 | 1 | 2 |
| Litjens[35] | 2016 | 1 | 2 | 1 | 1 | 1 | 1 | 1 |
| Ma[128] | 2020 | 3 | 2 | 1 | 3 | 3 | 1 | 2 |
| Menon[142] | 2022 | 2 | 2 | 3 | 3 | 3 | 3 | 3 |
| Mishra[144] | 2017 | 3 | 3 | 1 | 3 | 3 | 3 | 2 |
| Miyoshi[73] | 2020 | 3 | 2 | 1 | 1 | 3 | 2 | 1 |
| Mohlman[72] | 2020 | 3 | 2 | 1 | 3 | 3 | 1 | 1 |
| Naito[71] | 2021 | 3 | 2 | 1 | 1 | 3 | 1 | 1 |
| Nasir-Moin[121] | 2021 | 2 | 1 | 1 | 1 | 3 | 1 | 1 |
| Noorbakhsh[95] | 2020 | 2 | 2 | 3 | 3 | 3 | 1 | 3 |
| Rasmussen[45] | 2020 | 1 | 1 | 1 | 3 | 1 | 1 | 2 |
| Roy[101] | 2021 | 2 | 2 | 3 | 3 | 3 | 3 | 3 |
| Sabol[125] | 2020 | 2 | 3 | 1 | 3 | 3 | 3 | 2 |
| Sadeghi[102] | 2019 | 2 | 2 | 1 | 3 | 3 | 3 | 2 |
| Sali[119] | 2020 | 3 | 2 | 3 | 1 | 3 | 1 | 3 |
| Schau[70] | 2020 | 3 | 2 | 1 | 3 | 3 | 1 | 1 |
| Schilling[143] | 2018 | 3 | 2 | 3 | 1 | 3 | 1 | 3 |
| Schrammen[126] | 2022 | 3 | 3 | 3 | 3 | 2 | 3 | 3 |
| Shin[69] | 2020 | 2 | 3 | 1 | 3 | 3 | 1 | 2 |
| Song[68] | 2013 | 3 | 2 | 3 | 3 | 3 | 3 | 3 |
| Song[44] | 2020a | 3 | 1 | 3 | 3 | 3 | 1 | 3 |
| Song[93] | 2020b | 1 | 1 | 1 | 1 | 1 | 1 | 1 |
| Steinbuss[43] | 2020 | 3 | 2 | 1 | 3 | 3 | 1 | 1 |
| Steiner[103] | 2018 | 3 | 1 | 1 | 1 | 3 | 1 | 1 |
| Sun[67] | 2020 | 3 | 2 | 1 | 1 | 3 | 3 | 1 |
| Swiderska-Chadaj[58] | 2020 | 3 | 1 | 1 | 3 | 3 | 1 | 2 |
| Syed[120] | 2021 | 3 | 3 | 1 | 2 | 3 | 1 | 1 |
| Syrykh[137] | 2020 | 3 | 2 | 1 | 3 | 3 | 1 | 1 |
| Tabibu[134] | 2019 | 2 | 2 | 3 | 3 | 3 | 1 | 3 |
| Tsuneki[42] | 2021 | 2 | 2 | 1 | 3 | 2 | 1 | 1 |
| Tsuneki[57] | 2022 | 1 | 1 | 1 | 3 | 1 | 1 | 1 |
| Tung[41] | 2022 | 2 | 2 | 1 | 1 | 3 | 1 | 3 |
| Uegami[112] | 2022 | 1 | 2 | 1 | 1 | 1 | 1 | 1 |
| Valkonen[104] | 2017 | 3 | 2 | 3 | 3 | 3 | 3 | 3 |
| Wang KS[48] | 2021 | 3 | 1 | 1 | 3 | 3 | 1 | 3 |
| Wang L[66] | 2020 | 3 | 1 | 1 | 2 | 3 | 1 | 1 |
| Wang Q[50] | 2021 | 2 | 1 | 3 | 3 | 3 | 1 | 3 |
| Wang S[39] | 2019 | 3 | 2 | 3 | 3 | 3 | 1 | 3 |
| Wang X[65] | 2020 | 2 | 1 | 3 | 3 | 3 | 1 | 3 |
| Wang C[40] | 2017 | 2 | 2 | 3 | 3 | 3 | 3 | 3 |
| Wei[122] | 2020 | 3 | 1 | 1 | 1 | 1 | 1 | 1 |
| Wei[109] | 2019 | 3 | 2 | 1 | 1 | 3 | 1 | 1 |
| Wu[49] | 2020 | 3 | 3 | 3 | 3 | 3 | 3 | 3 |
| Xu[140] | 2017 | 2 | 2 | 1 | 3 | 3 | 1 | 3 |
| Xu[38] | 2021 | 2 | 1 | 1 | 3 | 2 | 1 | 3 |
| Yan[37] | 2022 | 2 | 3 | 3 | 3 | 3 | 3 | 3 |
| Yang[110] | 2021 | 3 | 1 | 1 | 3 | 3 | 1 | 1 |
| Yang[118] | 2022 | 3 | 3 | 1 | 3 | 3 | 3 | 2 |
| Yu[136] | 2020 | 2 | 2 | 1 | 3 | 3 | 1 | 1 |
| Yu[138] | 2019 | 2 | 3 | 1 | 3 | 3 | 3 | 1 |
| Yu[139] | 2021 | 3 | 3 | 1 | 3 | 3 | 1 | 1 |
| Zhang[64] | 2022 | 3 | 3 | 1 | 1 | 2 | 2 | 1 |
| Zhao[63] | 2021 | 2 | 2 | 3 | 3 | 3 | 1 | 2 |
| Zheng[111] | 2022 | 3 | 2 | 3 | 3 | 3 | 2 | 2 |
| Zhou[127] | 2021 | 3 | 1 | 1 | 3 | 3 | 3 | 1 |
| Zhu[56] | 2021 | 3 | 1 | 1 | 3 | 3 | 1 | 1 |

## S5 – Other accuracy / performance metrics for papers not included in the meta-analysis

| First author | Publication year | Reported performance (indication of best performance where multiple sets of results) |
|---|---|---|
| Abdeltawab[133] | 2021 | Average accuracy 0.957; sensitivity 0.920; specificity 0.971 |
| Alheejawi[94] | 2021 | Accuracy 97.7%, precisions 83.22, recall 87.08%, dice 85.10, Jaccard 74.07 |
| Ba[129] | 2021 | Overal accuracy 0.867. Best chronic atrophic gastritis: AUC 0.91, sens 0.952, spec 0.992, accuracy 0.986. Values given per disease class: Sensitivity 0.790-0.985; Specificity 0.829-1.000. |
| BenTaieb[135] | 2017 | Best accuracy Proposed model at 3 levels: 90.0% |
| Chen[105] | 2021 | (Best ADC & SCC) ADC AUC 0.9594 (0.9500-0.9689); SCC AUC 0.9414 (0.9243-0.9593) |
| Chen[116] | 2020 | (Detecting liver cancer) accuracy 0.960; Precision 0.945; Recall 1.000; F1 score 0.971. 89.6% accuracy for grade prediction |
| Chen[106] | 2022 | AUC 0.984 (per slide accuracy tumour detection WIFPS); accuracy 0.903; sensitivity 0.868; specificity 0.946 |
| Coudray[107] | 2018 | Normal vs tumour AUC 0.993 (0.974-1.0); 3 class at 5x best AUC 0.981 (0.968-0.980) |
| Cruz-Roa[96] | 2018 | Dice 0.76 +/- 0.20; PPV 0.72 +/- 0.22; NPV 0.97 +/- 0.05. (TPR 87%, TNR 92%, FPR 8%, FNR 13 |
| Cruz-Roa[97] | 2017 | Dice 0.7586 +/- 0.2006; PPV 0.7162 +/- 0.2204; NPV 0.9677 +/- 0.0511 |
| Dehkharghanian[108] | 2021 | Precision 0.92; Recall 0.91; F1 score 0.91 (average), accuracy 0.86-0.91 |
| del Amor[115] | 2021 | Sensitivity 0.9285; Specificity 0.9202; PPV 0.8622; NPV 0.9599; F1 score 0.8942; Accuracy 0.9231; AUC 0.9244. |
| DiPalma[141] | 2021 | KD (ADv2 model) - Coeliac: Accuracy 87.2, F1 score 75.86, Precision 76.46, Recall 78.0; KD model - Lung: Accuracy 94.18, F1 score 79.63, Precision 79.75, Recall 82.0; KD model - Renal: Accuracy 89.11, F1 Score 77.1, Precision 75.66, Recall 82.64. |
| Duran-Lopez[130] | 2021 | Accuracy 94.24%; Sensitivity 98.87%; Precision 90.23%; F1 score 94.33%; AUC 0.94 |
| Feng[123] | 2021 | Segmentation task: DSC 77.89%, Classification task: AUC 100% |
| Han[131] | 2020 | AlexNet-TCM: AUROC 0.964; error rate 6.1%; FNR 15.1%; FPR 5.8% |
| Haryanto[124] | 2021 | Best model taken as 300px+50px overlap. For image classification as malignant: Warwick dataset: sensitivity: 0.69; specificity: 0.93. UI dataset: sensitivity: 0.98; specificity: 1 |
| Huang[132] | 2021 | Distinguishing cancer from benign epithelium & stroma: AUROC=0.92 (95%CI 0.88-0.95);<br>Cancer detection: weighted k = 0.97 (95%CI 0.96-0.98);<br>Cancer grading: weighted k = 0.98 (95%CI 0.96-1) |
| Johny[98] | 2021 | Accuracy 0.9184; Precision 0.9185; Recall 0.9183; F1 score 0.9183; AUC 0.97 (triangular model) |
| Khalil[99] | 2022 | Precision 0.892; Recall 0.837; F1 score 0.844; mIoU 0.749 |
| Kiani[117] | 2020 | Accuracy 0.885 (0.710-0.960) (CNN alone on internal set); Accuracy 0.842 (0.808-0.876) (CNN alone on external set) |
| Kimeswenger[113] | 2020 | Accuracy 0.95; F1 score 0.97; AUC 0.99; Sensitivity 0.96; Specificity 0.93. |
| Li[114] | 2021 | AUC 0.971 |
| Lin[100] | 2019 | FROC (tumour localisation): 0.8533; AUROC (classification): 0.9875. |
| Ma[128] | 2020 | AUC 0.9876; accuracy 96%; specificity 93.3%; sensitivity 98.7% |
| Menon[142] | 2022 | Accuracy: BRCA 0.97, COAD 0.99, KICH 0.98, KIRP 0.95, LIHC 0.98; LUAD 0.95, LUSC 0.95, PRAD 0.92, READ 0.97, STAD 0.96 |
| Mishra[144] | 2017 | Accuracy 0.924; Precision 0.97; Recall 0.94; F1-score 0.95 |
| Nasir-Moin[121] | 2021 | Accuracy model + pathologist best: 80.8% (78.8-82.8) |
| Noorbakhsh[95] | 2020 | All tumour types (19) slide level: AUC 0.995 (+/- 0.008). All tumours types tile based: accuracy 0.91 (+/- 0.05); precision 0.97 (+/- 0.02); recall 0.90 (+/- 0.06); specificity 0.86 (+/- 0.07) |
| Roy[101] | 2021 | Accuracy 0.922; Precision 0.931; Recall 0.887; F1 score 0.908. |
| Sabol[125] | 2020 | CNN Balanced: Accuracy 92.74%; Precision 92.5%; Recall 92.76%; F1 92.64% |
| Sadeghi[102] | 2019 | 97.8% accuracy on validation set. On testing the 25% quantile of the probability score of the predictions increased from 0.48 to 0.89, and the median of the data increased from 0.95 to 0.99. |
| Sali[119] | 2020 | Best model GMM-RF: Average - accuracy 0.952 (0.915-0.989); AUC 0.986 (0.970-1.000); Precision 0.9555 (0.930-0.980); Recall 0.941 (0.903-0.979); F1 score 0.942 (0.904-0.981) |
| Schilling[143] | 2018 | Sensitivity 87.5%; Specificity 80%; PPV 83%; F1 score 88.9%; NPV 100% |
| Schrammen[126] | 2022 | AUROC 0.980 (0.975, 0.984) (on training set) |
| Song[44] | 2020 | Accuracy 90.4%; AUC 0.92; |
| Steiner[103] | 2018 | Sensitivity 91.2% (86-96.5%) P=0.023 (assisted read across images on case basis); AUC 98.5-0.99 |
| Syed[120] | 2021 | Multi-zoom ResNet50 patch level (same CM): Macro AUC 0.95; Accuracy 95% at patch level, sensitivity 0.96, specificity 0.97, PPV 0.96, NPV 0.97, Precision 0.94, Recall 0.94, F1 score 0.94. Modified ReNet50 with the ensemble: AUC 0.99, Accuracy 98.3%, Sensitivity 95%, Specificity 96%. Multi-zoom ResNet50 biopsy level: AUC 0.99; accuracy 0.98; sensitivity 0.96; Specificity 0.97; PPV 0.96; NPV 0.97; Precision 0.94; Recall 0.94. |
| Syrykh[137] | 2020 | AUC 0.99, accuracy 91% |
| Tabibu[134] | 2019 | ResNet-18 (KIRC) Cancer v Normal: patch wise accuracy 93.39; Precision 93.41; Recall 92.95; Slide wise AUC 0.99. |
| Uegami[112] | 2022 | Test set: Best AUC 0.88 (0.78-0.98). Sensitivity 0.89; Specificity 0.74. |
| Valkonen[104] | 2017 | Training: Accuracy 93%; Sensitivity 92.6%; Specificity 93.3%; F-score 0.93. Best AUC 0.98464 (0.97995 - 0.98932) cross validation. Random Forest sensitivity 92.6%, specificity 93.3%, F-score 0.93. |
| Wei[122] | 2020 | Internal mean: accuracy 93.5%; sensitivity 86.8%; specificity 95.7%. External mean: accuracy 87.0%; sensitivity 77.7%; specificity 91.6% |
| Wei[109] | 2019 | Kappa score 0.525; average agreement 66.6%; robust agreement 76.7% |
| Xu[140] | 2017 | Brain cancer classification (best): Accuracy 97.8%. Segmenting: accuracy 84%. CRC binary best: accuracy 98.0%. CRC multiclass 87.2%. |
| Yang[110] | 2021 | EfficientNetB5 on SYSU1 (best): Macro average AUC 0.988 (0.982-0.994); accuracy 0.860; weighted F1 score 0.860 |
| Yang[118] | 2022 | (FA-MSCN 5x_2.5x) Sensitivity 0.96; Intersection over union (IOU) 0.89 |
| Yu[136] | 2020 | AUC 0.975 (+/- 0.001) (Tumour detection) |
| Yu[138] | 2019 | AUC 0.985 (+/- 0.004) (SCC vs benign); AUC 0.971 (+/- 0.007) (AdenoCa vs benign) |
| Yu[139] | 2021 | AUC 0.996 (CI 0.949-0.984) (case level but 1 slide per case) |
| Zheng[111] | 2022 | TCGA ext test set normal v tumour: AUC 0.980 (+/- 0.04). 3 label task TCGA: Average accuracy 82.3; average AUC 92.8. |
| Zhou[127] | 2021 | Combination framework: Accuracy 0.946, Precision 0.964, Recall 0.982, F1 score 0.973 |

## S6 – Meta analysis: additional data & source of data

| Author | TP | FN | FP | TN | N |
|---|---|---|---|---|---|
| Aatresh 2021[82] | 90 | 3 | 0 | 50 | 143 |
| Achi 2019[81] | 176 | 4 | 4 | 56 | 240 |
| Ashraf 2022[47] | 485 | 9 | 16 | 231 | 741 |
| Cengzig 2022[55] | 63322 | 9029 | 5848 | 23507 | 101706 |
| Cho 2019[46] | 25 | 0 | 0 | 25 | 50 |
| Choudhary 2021[54] | 56333 | 3311 | 3289 | 20325 | 83258 |
| da Silva 2021[62] | 173 | 2 | 27 | 377 | 579 |
| De Logu 2020[80] | 1074 | 48 | 18 | 773 | 1913 |
| Esteban 2019[61] | 16 | 2 | 0 | 1 | 19 |
| Fenstermaker 2020[60] | 12046 | 0 | 85 | 3000 | 15131 |
| Fu 2021[79] | 35 | 0 | 0 | 12 | 47 |
| Hameed 2020[53] | 86 | 2 | 6 | 76 | 170 |
| Han 2020[59] | 32092 | 5689 | 70530 | 1140166 | 1248477 |
| Hekler 2019[78] | 38 | 12 | 20 | 30 | 100 |
| Hohn 2021*[77] | 60 | 5 | 17.4 | 49.6 | 132 |
| Iizuka (Colon) 2020[36] | 21 | 0 | 33 | 446 | 500 |
| Iizuka (Gastro) 2020[36] | 56 | 5 | 23 | 416 | 500 |
| Jin 2020[52] | 13435 | 2949 | 1999 | 14385 | 32768 |
| Kanavati 2020[76] | 586 | 5 | 41 | 48 | 680 |
| Kanavati 2021[51] | 431 | 127 | 30 | 794 | 1382 |
| Li 2018[75] | 6944 | 2 | 56 | 998 | 8000 |
| Li 2020[74] | 171 | 3 | 24 | 61 | 259 |
| Litjens (Prostate) 2016[35] | 43 | 2 | 0 | 30 | 75 |
| Litjens (Breast) 2016[35] | 16 | 2 | 16 | 40 | 74 |
| Miyoshi 2020[73] | 78 | 1 | 2 | 19 | 100 |
| Mohlman 2020[72] | 741 | 101 | 860 | 2372 | 4074 |
| Naito 2021[71] | 80 | 6 | 1 | 33 | 120 |
| Rasmussen 2020[45] | 446 | 15 | 2 | 508 | 971 |
| Schau 2020[70] | 16250 | 4737 | 4862 | 5228 | 31077 |
| Shin 2020[69] | 594 | 26 | 212 | 408 | 1240 |
| Song 2013[68] | 69 | 13 | 11 | 67 | 160 |
| Song 2020[93] | 630 | 3 | 405 | 2174 | 3212 |
| Steinbuss 2020[43] | 16 | 8 | 8 | 76 | 108 |
| Sun 2020[67] | 46 | 13 | 0 | 141 | 200 |
| Swiderska Chadaj 2020[58] | 55 | 3 | 4 | 23 | 85 |
| Tsuneki 2021[42] | 63 | 11 | 153 | 1572 | 1799 |
| Tsuneki 2022[57] | 695 | 38 | 1 | 33 | 767 |
| Tung 2022[41] | 157 | 28 | 22 | 343 | 550 |
| Wang KS 2021[48] | 3940 | 48 | 9 | 1842 | 5839 |
| Wang L 2020[66] | 60289 | 5963 | 1215 | 15660 | 83127 |
| Wang Q 2021[50] | 38 | 4 | 3 | 65 | 110 |
| Wang S 2019[39] | 104 | 2 | 18 | 76 | 200 |
| Wang X 2020[65] | 170 | 0 | 1 | 14 | 185 |
| Wang C 2017[40] | 116 | 9 | 10 | 865 | 1000 |
| Wu 2020*[49] | 17.6 | 18.4 | 15.7 | 376.3 | 428 |
| Xu 2021[38] | 19300 | 700 | 360 | 19640 | 40000 |
| Yan 2022[37] | 1397 | 14 | 267 | 8322 | 10000 |
| Zhang 2022[64] | 1056 | 26 | 34 | 558 | 1674 |
| Zhao 2021[63] | 213 | 13 | 21 | 82 | 329 |
| Zhu 2021[56] | 904 | 4 | 0 | 9 | 917 |

*Data provided by authors as averages of a cross validation (not whole numbers)

| Colour key for source of meta-analysis data | |
|---|---|
| | Retrieved from study / supplementary materials |
| | Multiclass confusion matrix in study reduced to 2x2 table |
| | Back-calculated from data provided in study |
| | Provided by author |
| | Back-calculated from data provided by author |
| | Multiclass confusion matrix provided by author reduced to 2x2 table |

# S7 – Raw data for forest plots Figure 4 (main text)

| Author | Sensitivity | Lower 95% CI | Upper 95% CI |
|---|---|---|---|
| Aatresh 2021[82] | 0.97 | 0.91 | 0.99 |
| Achi 2019[81] | 0.98 | 0.94 | 0.99 |
| Ashraf 2022[47] | 0.98 | 0.97 | 0.99 |
| Cengzig 2022[55] | 0.88 | 0.87 | 0.88 |
| Cho 2019[46] | 1.00 | 0.87 | 1.00 |
| Choudhary 2021[54] | 0.94 | 0.94 | 0.95 |
| da Silva 2021[62] | 0.99 | 0.96 | 1.00 |
| De Logu 2020[80] | 0.96 | 0.94 | 0.97 |
| Esteban 2019[61] | 0.89 | 0.67 | 0.97 |
| Fenstermaker 2020[60] | 1.00 | 1.00 | 1.00 |
| Fu 2021[79] | 1.00 | 0.90 | 1.00 |
| Hameed 2020[53] | 0.98 | 0.92 | 0.99 |
| Han 2020[59] | 0.85 | 0.85 | 0.85 |
| Hekler 2019[78] | 0.76 | 0.63 | 0.86 |
| Hohn 2021[77] | 0.92 | 0.83 | 0.97 |
| Iizuka (Colon) 2020[36] | 1.00 | 0.85 | 1.00 |
| Iizuka (Gastric) 2020[36] | 0.92 | 0.82 | 0.96 |
| Jin 2020[52] | 0.82 | 0.81 | 0.83 |
| Kanavati 2020[76] | 0.99 | 0.98 | 1.00 |
| Kanavati 2021[51] | 0.77 | 0.74 | 0.81 |
| Li 2018[75] | 1.00 | 1.00 | 1.00 |
| Li 2020[74] | 0.98 | 0.95 | 0.99 |
| Litjens (Breast) 2016[35] | 0.89 | 0.67 | 0.97 |
| Litjens (Prostate) 2016[35] | 0.96 | 0.85 | 0.99 |
| Miyoshi 2020[73] | 0.99 | 0.93 | 1.00 |
| Mohlman 2020[72] | 0.88 | 0.86 | 0.90 |
| Naito 2021[71] | 0.93 | 0.86 | 0.97 |
| Ramussen 2020[45] | 0.97 | 0.95 | 0.98 |
| Schau 2020[70] | 0.77 | 0.77 | 0.78 |
| Shin 2020[69] | 0.96 | 0.94 | 0.97 |
| Song 2013[68] | 0.84 | 0.75 | 0.90 |
| Song 2020[93] | 1.00 | 0.99 | 1.00 |
| Steinbuss 2020[43] | 0.67 | 0.47 | 0.82 |
| Sun 2020[67] | 0.78 | 0.66 | 0.87 |
| Swiderska Chadaj 2020[58] | 0.95 | 0.86 | 0.98 |
| Tsuneki 2021[42] | 0.85 | 0.75 | 0.91 |
| Tsuneki 2022 | 0.95 | 0.93 | 0.96 |
| Tung 2022[41] | 0.85 | 0.79 | 0.89 |
| Wang C 2017[40] | 0.93 | 0.87 | 0.96 |
| Wang KS 2021[48] | 0.99 | 0.98 | 0.99 |
| Wang L 2019[66] | 0.91 | 0.91 | 0.91 |
| Wang Q 2021[50] | 0.90 | 0.78 | 0.96 |
| Wang S 2019[39] | 0.98 | 0.93 | 0.99 |
| Wang X 2020[65] | 1.00 | 0.98 | 1.00 |
| Wu 2020[49] | 0.49 | 0.33 | 0.65 |
| Xu 2021[38] | 0.96 | 0.96 | 0.97 |
| Yan 2022[37] | 0.99 | 0.98 | 0.99 |
| Zhang 2022[64] | 0.98 | 0.97 | 0.98 |
| Zhao 2021[63] | 0.94 | 0.90 | 0.97 |
| Zhu 2021[56] | 1.00 | 1.00 | 1.00 |

| Author | Specificity | Lower 95% CI | Upper 95% CI |
|---|---|---|---|
| Aatresh 2021[82] | 1.00 | 0.93 | 1.00 |
| Achi 2019[81] | 0.93 | 0.84 | 0.97 |
| Ashraf 2022[47] | 0.94 | 0.90 | 0.96 |
| Cengzig 2022[55] | 0.80 | 0.80 | 0.81 |
| Cho 2019[46] | 1.00 | 0.87 | 1.00 |
| Choudhary 2021[54] | 0.86 | 0.86 | 0.87 |
| da Silva 2021[62] | 0.93 | 0.90 | 0.95 |
| De Logu 2020[80] | 0.98 | 0.96 | 0.99 |
| Esteban 2019[61] | 1.00 | 0.21 | 1.00 |
| Fenstermaker 2020[60] | 0.97 | 0.97 | 0.98 |
| Fu 2021[79] | 1.00 | 0.76 | 1.00 |
| Hameed 2020[53] | 0.93 | 0.85 | 0.97 |
| Han 2020[59] | 0.94 | 0.94 | 0.94 |
| Hekler 2019[78] | 0.60 | 0.46 | 0.72 |
| Hohn 2021[77] | 0.74 | 0.62 | 0.83 |
| Iizuka (Colon) 2020[36] | 0.93 | 0.90 | 0.95 |
| Iizuka (Gastric) 2020[36] | 0.95 | 0.92 | 0.96 |
| Jin 2020[52] | 0.88 | 0.87 | 0.88 |
| Kanavati 2020[76] | 0.54 | 0.44 | 0.64 |
| Kanavati 2021[51] | 0.96 | 0.95 | 0.97 |
| Li 2018[75] | 0.95 | 0.93 | 0.96 |
| Li 2020[74] | 0.72 | 0.61 | 0.80 |
| Litjens (Breast) 2016[35] | 0.71 | 0.59 | 0.82 |
| Litjens (Prostate) 2016[35] | 1.00 | 0.89 | 1.00 |
| Miyoshi 2020[73] | 0.91 | 0.71 | 0.97 |
| Mohlman 2020[72] | 0.73 | 0.72 | 0.75 |
| Naito 2021[71] | 0.97 | 0.85 | 0.99 |
| Rasmussen 2020[45] | 1.00 | 0.99 | 1.00 |
| Schau 2020[70] | 0.52 | 0.51 | 0.53 |
| Shin 2020[69] | 0.66 | 0.62 | 0.69 |
| Song 2013[68] | 0.86 | 0.76 | 0.92 |
| Song 2020[93] | 0.84 | 0.83 | 0.86 |
| Steinbuss 2020[43] | 0.90 | 0.82 | 0.95 |
| Sun 2020[67] | 1.00 | 0.97 | 1.00 |
| Swiderska Chadaj 2020[58] | 0.85 | 0.68 | 0.94 |
| Tsuneki 2021[42] | 0.91 | 0.90 | 0.92 |
| Tsuneki 2022 | 0.97 | 0.85 | 0.99 |
| Tung 2022[41] | 0.94 | 0.91 | 0.96 |
| Wang C 2017[40] | 0.99 | 0.98 | 0.99 |
| Wang KS 2021[48] | 1.00 | 0.99 | 1.00 |
| Wang L 2019[66] | 0.93 | 0.92 | 0.93 |
| Wang Q 2021[50] | 0.96 | 0.88 | 0.98 |
| Wang S 2019[39] | 0.81 | 0.72 | 0.88 |
| Wang X 2020[65] | 0.93 | 0.70 | 0.99 |
| Wu 2020[49] | 0.96 | 0.94 | 0.98 |
| Xu 2021[38] | 0.98 | 0.98 | 0.98 |
| Yan 2022[37] | 0.97 | 0.97 | 0.97 |
| Zhang 2022[64] | 0.94 | 0.92 | 0.96 |
| Zhao 2021[63] | 0.80 | 0.71 | 0.86 |
| Zhu 2021[56] | 1.00 | 0.70 | 1.00 |

## S8 – Supplementary forest plots of sensitivity and specificity for subgroups

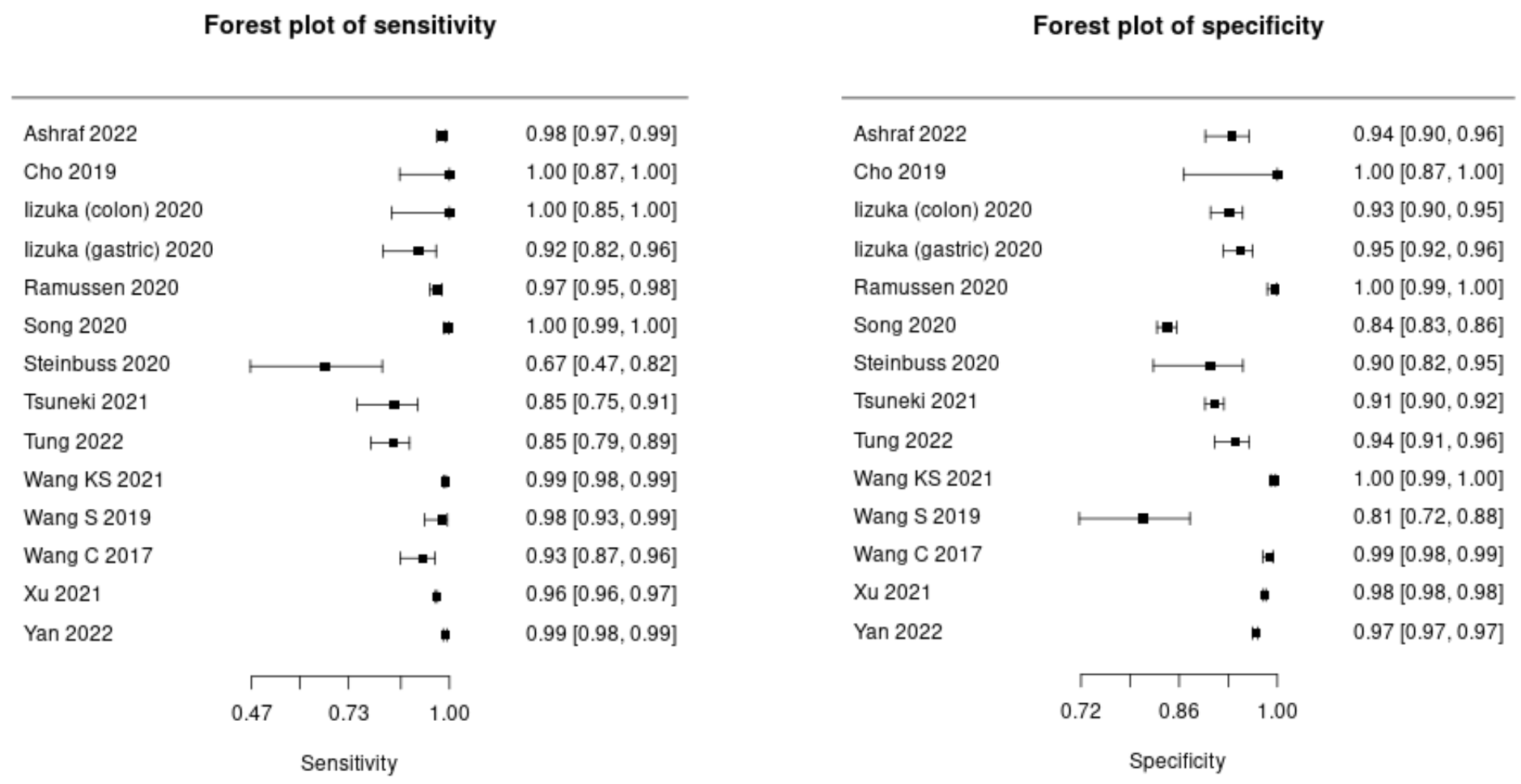

*S8a – Forest plots for sensitivity and specificity in studies of gastrointestinal pathology*

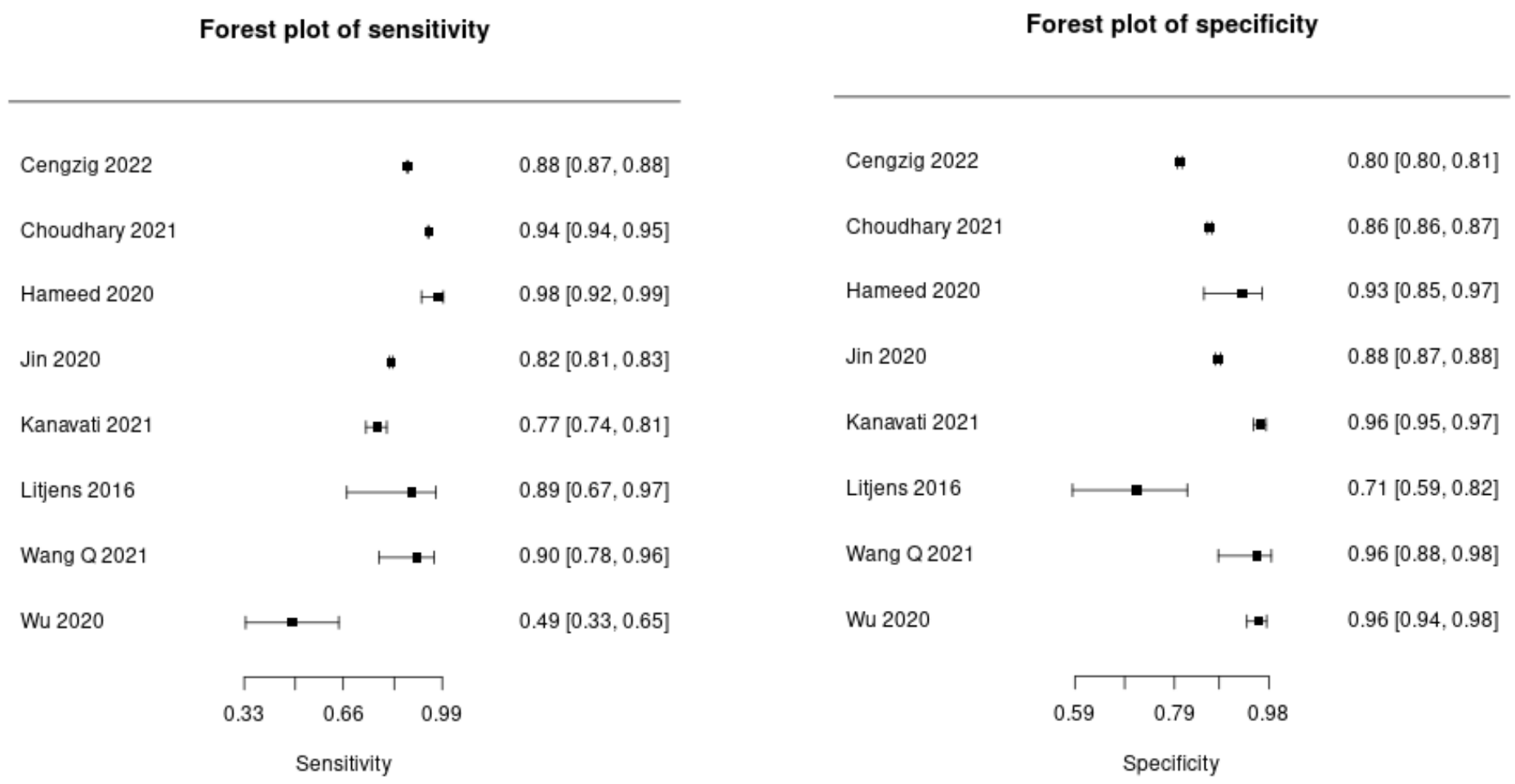

*S8b – Forest plots for sensitivity and specificity in studies of breast pathology*

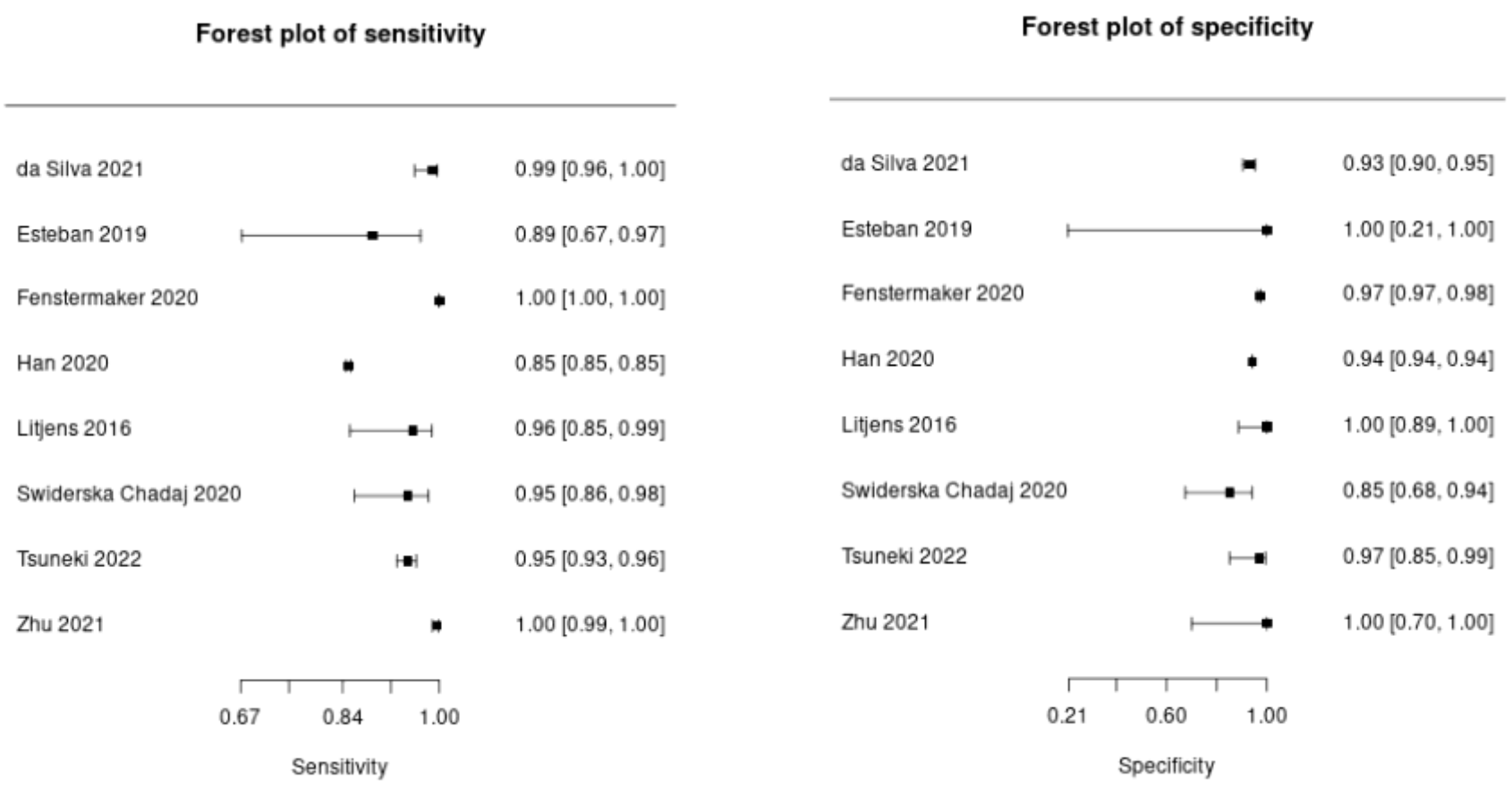

*S8c – Forest plots for sensitivity and specificity in studies of urological pathology*

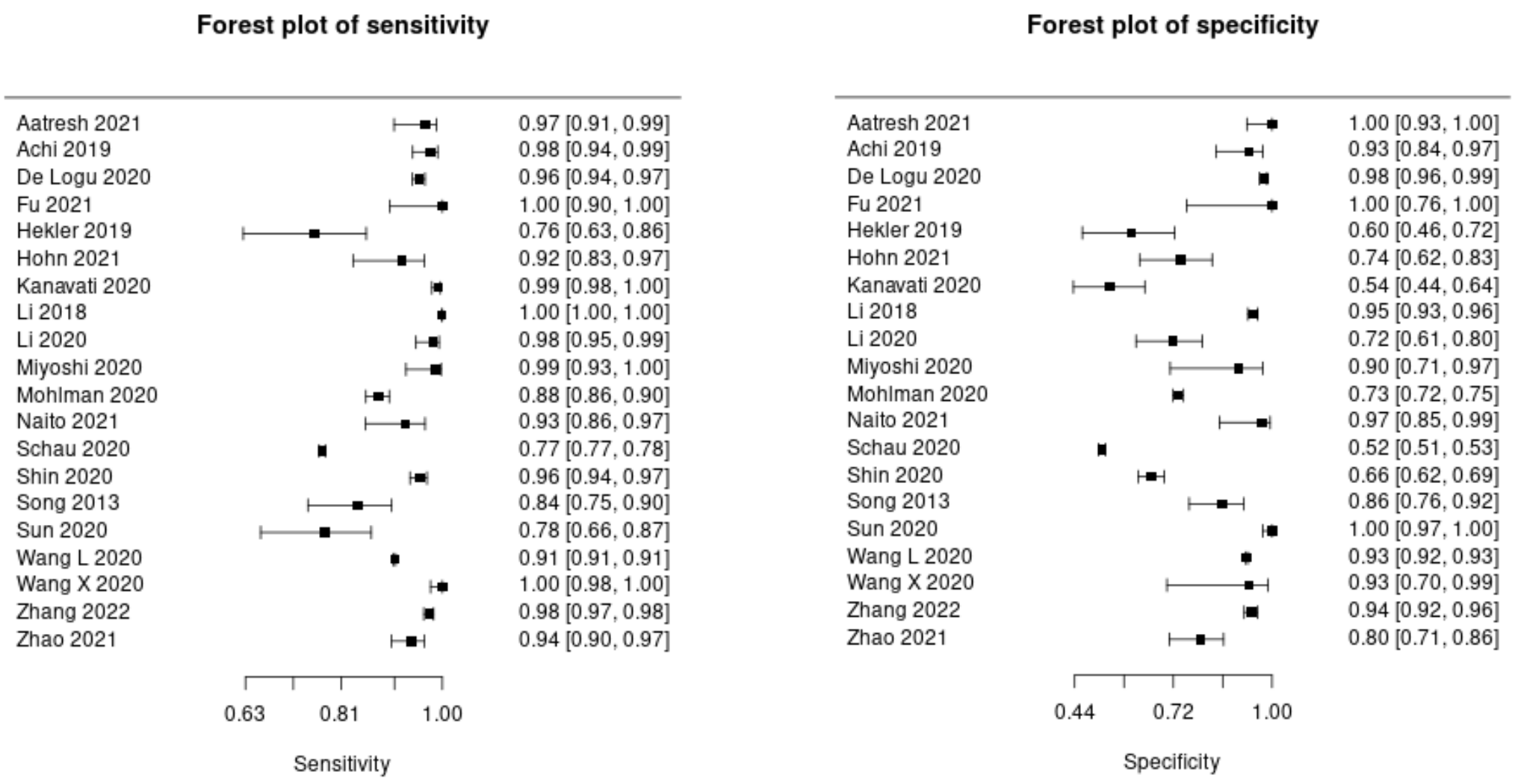

*S8d – Forest plots for sensitivity and specificity in studies of other pathologies*

## S9 – Performance by number of included data sources in the meta-analysis

| No. data sources | No. models | Mean sensitivity (%) | Mean specificity (%) |
|---|---|---|---|
| 1 | 23 | 89% | 88% |
| 2 | 18 | 95% | 92% |
| 3 | 4 | 93% | 92% |
| 4 | 1 | 99% | 54% |
| 5 | 1 | 85% | 91% |
| 6 | 1 | 95% | 97% |
| 14 | 1 | 99% | 100% |
| Not stated | 1 | 88% | 80% |

## S10 – Performance of models including an external validation in the meta-analysis

| External validation of the model | No. models | Mean sensitivity (%) | Mean specificity (%) |
|---|---|---|---|
| No | 26 | 91% | 87% |
| Unclear | 3 | 78% | 90% |
| Yes | 21 | 95% | 92% |

## S11 – Performance of models by unit of analysis in the meta-analysis

| Unit of analysis | No. models | Mean sensitivity (%) | Mean specificity (%) |
|---|---|---|---|
| Other | 2 | 74% | 95% |
| Patch / Tile | 28 | 91% | 90% |
| Slide | 20 | 95% | 88% |

## S12 – Further details of study characteristics for al included studies

| First author | Publication year | Funding source of research | Intended use* | Pathological subspecialty/ies | Total number of slides in study (other units if not provided) | Number of data sources | Is the dataset(s) open source | Is the test set independent of the training set |
|---|---|---|---|---|---|---|---|---|
| Aatresh[82] | 2021 | Science Engineering and Research Board, Department of Science and Technology, Govt. of India | Classifying subtypes of liver cancer | Hepatobiliary pathology | 398 WSI (141 WSI into 705 patches for TCGA), (257 WSI into 1338 patches for KMC) | 2 | Mixed | Unclear |
| Abdeltawab[133] | 2021 | No funder declared | Classifying subtypes of renal cancer | Uropathology | 64 WSIs | 1 | No | Yes |
| Achi[81] | 2019 | No funder declared | Classifying lymphoma subtypes | Haematopathology | 128 WSIs (equalling 2560 40x40 pixel patches) | 2 | Unclear | Yes |
| Alheejawi[94] | 2021 | Natural Sciences and Engineering Research Council of Canada; Ministry of Higher Education and Scientific Research, Iraq; Imam Ja'afar Al Sadiq University, Iraq | Detecting melanoma | Dermatopathology | 4 WSIs | 1 | No | Yes |
| Ashraf[47] | 2022 | Seegene Medical Foundation, South Korea | Detecting gastric cancer | Gastrointestinal pathology | 905 WSIs and 327,680 96x96 pixel patches | 2 | Mixed | Yes |
| Ba[129] | 2021 | PLA General Hospital Medical Big Data and Artificial Intelligence Project | Classifying subtypes of gastritis | Gastrointestinal pathology | 1250 WSIs | 1 | No | Yes |
| BenTaieb[135] | 2017 | Natural Sciences and Engineering Research Council of Canada | Classifying subtypes of ovarian cancer | Gynaecological pathology | 133 WSIs | 1 | Yes | Yes |
| Cengzig[55] | 2022 | No funder declared | Detecting breast cancer | Breast pathology | 398,381 50x50 size patches | Not stated | Unclear | Unclear |
| Chen[105] | 2021 | Ministry of Sciences and Technology Taiwan | Classifying subtypes of lung cancer | Cardiothoracic pathology | 7003 WSIs hospitals set; 1044 WSIs TCGA test set. | 4 | Mixed | Yes |
| Chen[116] | 2020 | Opening Fund of Engineering Research Center of Cognitive Healthcare of Zhejiang Province, Zhejiang Medical Health Science and Technology Project, National Natural Science Foundation of China | Detecting liver cancer; grading liver cancer severity | Hepatobiliary pathology | 592 WSIs | 2 | Mixed | Yes |
| Chen[106] | 2022 | National Key R&D program of China; National Natural Science Foundation of China; Guangdong Natural Science Foundation. | Detecting lung cancer, classifying subtype of lung cancer | Cardiothoracic pathology | 1914 cases | 3 | No | Yes |
| Cho[46] | 2019 | National Research Foundation of Korea; Catholic Medical Centre Research Foundation | Detecting gastric cancer | Gastrointestinal pathology | 803 WSIs | 2 | Mixed | Yes |
| Choudhary[54] | 2021 | No funder declared | Detecting breast cancer | Breast pathology | 162 WSIs | 1 | Yes | Yes |
| Coudray[107] | 2018 | Cancer Centre Support Grant, Laura and Isaac Perlmutter Cancer Centre. | Detecting lung cancer & classification of non-small cell lung cancer subtypes | Cardiothoracic pathology | 1634 WSIs (TCGA) + 340 WSIs (New York) independent set | 2 | Mixed | Yes |
| Cruz-Roa[96] | 2018 | Administrative Department of Science, Technology and Innovation - Colciencias, Universidad Nacional de Colombia; Universidad de los Llanos; the National Cancer Institute of the National Institutes of Health; National Institute of Diabetes and Digestive and Kidney Diseases; National Center for Research Resources; United States Department of Defense Prostate Cancer Synergistic Idea Development Award; United States Department of Defense Lung Cancer Idea Development New Investigator Award; United States Department of Defense Prostate Cancer Idea Development Award; United States Department of Defense Peer Reviewed Cancer Research Program Case Comprehensive Cancer Center Pilot Grant; VelaSano Grant, Cleveland Clinic; the Wallace H. Coulter Foundation Program Case Western Reserve University. | Detecting breast cancer | Breast pathology | 945 cases | 4 | Mixed | Yes |
| Cruz-Roa[97] | 2017 | DGI-Unillanos; Administrative Department of Science, Technology and Innovation of Colombia; National Cancer Institutes of the National Institutes of Health; the National Institute of Diabetes and Digestive and Kidney diseases; National Center for Research Resources; DOD Prostate Cancer Synergistic Idea Development Award; DOD Lung Cancer Idea Development New Investigator Award; DOD Prostate Cancer Idea Development Award; DOD Peer Reviewed Cancer Research Program; Cleveland Clinic; Wallace H. Coulter Foundation Program, Case Western Reserve University. | Detecting breast cancer | Breast pathology | 605 patients | 4 | Mixed | Yes |
| da Silva[62] | 2021 | Paige; Breast Cancer Research Foundation; National Institutes of Health / National Cancer Institute; P50 grant; | Detecting prostate cancer | Uropathology | 661 WSIs (from 579 unique needle core biopsy parts | 1 | No | Yes |
| De Logu[80] | 2020 | Associazione Italiana per la Ricerca sul Cancro | Detecting melanoma | Dermatopathology | 100 WSIs | 3 | No | Yes |
| Dehkharghanian[108] | 2021 | Government of Ontario, Canada and the Ontario Research Fund-Research Excellence Gigapixel image identification consortium | Classifying lung cancer subtypes | Cardiothoracic pathology | 758 WSIs | 2 | Mixed | Yes |
| del Amor[115] | 2021 | Horizon 2020, the Spanish Ministry of Economy and Competitiveness, Instituto de | Detecting spitzoid melanocytic lesions | Dermatopathology | 53 WSIs | 1 | No | Yes |
| | | Salud Carlos III, GVA, Polytechnic University of Valencia, Marie Skłodowska Curie grant | | | | | | |
| DiPalma[141] | 2021 | US National Library of Medicine, US National Cancer Institute | Detecting coeliac disease; classifying lung cancer subtypes; classifying renal cancer subtypes | Multiple | Coeliac: 1364 patients; Lung: 269 WSIs; Renal 882 WSIs. | 2 | Mixed | Yes |
| Duran-Lopez[130] | 2021 | Spanish Agencia Estatal de Investigation, European Regional Development Fund | Detecting prostate cancer | Uropathology | 332 WSIs | 1 | No | Unclear |
| Esteban[61] | 2019 | Ministerio de Economía y Competitividad. | Detecting prostate cancer | Uropathology | 79 WSIs from SICAPv1; and ext set 593 patches for testing from Gertych et al | 1 | Mixed | Yes |
| Feng[123] | 2021 | National Key Research and Development Program of China; National Natural Science Foundation of China; Zhejiang University Education Foundation; Zhejiang public welfare technology research project; Key Laboratory of Medical Neurobiology of Zhejiang Province; NSF Grant. | Detecting colorectal cancer | Gastrointestinal pathology | 1000 WSIs | 1 | Yes | Yes |
| Fenstermaker[60] | 2020 | No funders declared | Detecting renal cell cancer. Classifying subtypes of RCC. | Uropathology | 42 patients | 1 | Yes | Yes |
| Fu[79] | 2021 | Foundation of Beijing Municipal Science and Technology Commission; National Key Research and Development Program of China; National Natural Science Foundation of China. | Detecting pancreatic ductal adenocarcinoma | Hepatobiliary pathology | 283 WSIs | 2 | Mixed | Yes |
| Hameed[53] | 2020 | Basque Country project MIFLUDAN; eVida Research Group IT 905-16 (University of Deusto, Spain) | Detecting breast cancer | Breast pathology | 845 areas/patches from 544 WSIs. | 1 | No | Yes |
| Han[131] | 2020a | No funders declared | Detecting prostate cancer. | Uropathology | 299 WSIs | 1 | No | Yes |
| Han[59] | 2020b | Canadian Institute of Health Research; Ontario Institute for Cancer Research; Prostate Canada; Natural Sciences and Engineering Research Council of Canada | Detecting prostate cancer | Uropathology | 299 WSIs | 1 | No | Yes |
| Haryanto[124] | 2021 | Ministry of Research and Technology, Republic of Indonesia | Detecting colorectal cancer | Gastrointestinal pathology | 165 images + other images from University of Indonesia. (For best model (300px + 50px overlap), no. of CSW-generated images = 13,576 (2,984 (Warwick), 10,592 (UI)) | 2 | Mixed | Unclear |
| Hekler[78] | 2019 | No funders | Detecting melanoma | Dermatopathology | 695 WSIs from 595 patients | 1 | No | Yes |
| Hohn[77] | 2021 | Federal Ministry of Health, Berlin, Germany; Tumour Behaviour Prediction Initiative. | Detecting melanoma | Dermatopathology | 431 WSIs | 2 | No | Yes |
| Huang[132] | 2021 | PathomIQ | Detecting prostate cancer. | Uropathology | 1000 WSIs | 1 | No | Yes |
| Iizuka[36] | 2020 | No funders declared | Classifying gastric and colonic tumours | Gastrointestinal pathology | 10,186 WSIs | 2 | Mixed | Yes |
| Jin[52] | 2020 | CancerCare Manitoba Founation; Natural Sciences and Engineering Research Council of Canada; University of Manitoba; Manitoba Medical Services Foundation Allen Rouse Basic Science Career Development Research Award. | Detecting breast cancer metastases in lymph nodes | Breast pathology | 327,680 patches (PCaM), 438 images (second dataset), 100 patches from 10 WSIs (Warwick) | 3 | Yes | Yes |
| Johny[98] | 2021 | No funders declared | Detecting breast cancer metastases in lymph nodes | Breast pathology | 327,680 patches from 400 WSIs | 1 | Yes | Yes |
| Kanavati[51] | 2021 | No funders declared | Detecting breast cancer and DCIS | Breast pathology | 3672 WSIs | 2 | No | Yes |
| Kanavati[76] | 2020 | Research Institute for Information Technology, Kyushu University | Detecting lung cancer | Cardiothoracic pathology | 5734 WSIs | 4 | Mixed | Yes |
| Khalil[99] | 2022 | Ministry of Science and Technology of Taiwan | Detecting breast cancer metastases in lymph nodes | Breast pathology | 188 WSIs (94 H&E, 94 matching IHC CK(AE1/AE3) WSIs) | 1 | No | Yes |
| Kiani[117] | 2020 | Department of Pathology (Stanford University) Stanford Machine Learning Group and the Stanford Center for Artificial Intelligence in Medicine & Imaging | Classification of liver tumour subtypes | Hepatobiliary pathology | 150 WSIs | 2 | Mixed | Yes |
| Kimeswenger[113] | 2020 | ERC; REA; Promedica Stiftung; Swiss Cancer Research Foundation; Clinical Research Priority Program (CRPP), University of Zurich; Swiss National Science Foundation; European Academic of Dermatology and Venereology. | Detecting basal cell carcinoma | Dermatopathology | 820 WSIs | 2 | No | Yes |
| Li[114] | 2021 | The National Key Research and Development Program of China; Natural Science Foundation of China; Hunan Province Science Foundation; Changsha Muncipal Natural Science Foundation; Scientific Research Fund of Hunan Provincial Education Department. | Detecting melanoma | Dermatopathology | 701 WSIs | 2 | Mixed | Yes |
| Li[75] | 2018 | No funder declared | Classifying subtypes of brain tumour | Neuropathology | 206 WSIs | 1 | No | Yes |
| Li[74] | 2020 | No funder declared | Detecting thyroid cancer | Head & neck pathology | 608 WSIs | 1 | No | Yes |
| Lin[100] | 2019 | Hong Kong Innovation and Technology Commission; Hong Kong Research Grants Council; Global Partnership Fund, University of Warwick. | Detect breast cancer metastases in lymph nodes | Breast pathology | 400 WSIs | 1 | Yes | Yes |
| Litjens[35] | 2016 | StITPro Foundation | Detecting breast cancer metastases in sentinel lymph nodes & prostate cancer grading | Multiple | Prostate: 225 WSIs; Breast: 271 WSIs. | 1 | No | Yes |
| Ma[128] | 2020 | Shanghai Science and Technology Committee; National Key R&D Program of China; National Natural Science Foundation of China; Cross-Institute Research Fund of Shanghai Jiao Tong University; Innovation Foundation of Translational Medicine of Shanghai Jiao Tong University School of Medicine; Technology Transfer Project of Science & Technology, Department of Shanghai Jiao Tong University School of Medicine | Detecting gastric cancer and classifying gastric disease | Gastrointestinal pathology | 763 WSIs | 1 | No | Yes |
| Menon[142] | 2022 | Ihub-Data, International Institute of Information and Technology, Hyderabad | Detect multiple cancer types | Multiple | 9792 WSIs | 1 | Yes | Yes |
| Mishra[144] | 2017 | Cancer Prevention and Research Institute of Texas (CPRIT) | Detecting osteosarcoma | Soft tissue & bone pathology | 82 WSIs (64,000 patches) | Unclear | No | Yes |
| Miyoshi[73] | 2020 | Chugai Pharmaceutical Co. Ltd | Classify subtypes of Lymphoma | Haematopathology | 388 sections | 1 | No | Yes |
| Mohlman[72] | 2020 | No funder declared | Classify subtypes of lymphoma | Haematopathology | 10,818 patches from unknown no. slides (70 cases) | 2 | No | Yes |
| Naito[71] | 2021 | Research Institute for Information Technology Kyushu University | Detecting pancreatic ductal adenocarcinoma | Hepatobiliary pathology | 532 WSIs | 1 | No | Yes |
| Nasir-Moin[121] | 2021 | National Cancer Institute; National Library of Medicine | Assisting the pathologist with classifying subtypes of colorectal polyp | Gastrointestinal pathology | 846 WSIs used in experiment + 60 WSIs for other purposes | 25 | No | Yes |
| Noorbakhsh[95] | 2020 | NIH Cloud Credits Model Pilot, NIH Big Data to Knowledge (BD2K) program; Google Cloud; NCI grant. | Detecting multiple cancer types and subtype classification | Multiple | 29,930 WSIs | 2 | Yes | Yes |
| Rasmussen[45] | 2020 | Nova Scotia Health Authority Research Fund | Detecting hereditary diffuse gastric cancer | Gastrointestinal pathology | 17,636 patches | 2 | No | Yes |
| Roy[101] | 2021 | No funders | Detecting invasive ductal carcinoma of the breast | Breast pathology | 162 WSIs; 277,524 patches | 1 | Yes | Unclear |
| Sabol[125] | 2020 | AI4EU project from European Union's Horizon 2020 research & innovation programme; Maria Currie RISE LIFEBOTS Exchange Grant; EU FlagEra Joint Progect Robocom++, 2017-2021 | Detecting colorectal cancer | Gastrointestinal pathology | 5000 tiles | 1 | Yes | Unclear |
| Sadeghi[102] | 2019 | BMBF grant | Detecting lymph node breast cancer metastases | Breast pathology | 500 WSI (camelyon 17) + 20,000 patches (cameylon 16) | 2 | Yes | Yes |
| Sali[119] | 2020 | National Institute of Diabetes and Digestive and Kidney Diseases of the National Institutes of Health. | Detecting dysplastic barretts oesophagus and non-dysplastic barretts oesophagus | Gastrointestinal pathology | 650 WSI | 1 | Unclear | Yes |
| Schau[70] | 2020 | National Cancer Institute; OHSU Center for Spatial Systems Biomedicine; Knight Diagnostic Laboratories; Biomedical Innovation Program Award, Oregon Clinical and Translational Research Institute. | Detecting liver metastasis and classifying origin site of liver metastases | Gastrointestinal pathology | 285 WSIs | 1 | Unclear | Yes |
| Schilling[143] | 2018 | No funder declared | Detecting Hirsprungs disease | Paediatric pathology | 307 WSIs | 1 | Yes | Yes |
| Schrammen[126] | 2022 | German Federal Ministry of Health; Max-Eder-Programme of the German Cancer Aid; NIHR; Yorkshire Cancer Research program; German Research Foundation; Interdisciplinary Research Program of the National Centre for Tumour Diseases, Germany; German Federal Ministry of Education and Research. | Detecting colorectal cancer | Gastrointestinal pathology | 3337 cases | 2 | No | Yes |
| Shin[69] | 2020 | Ministry of Trade, Industry & Energy (Korea); Ministry of Health & Welfare (Korea) | Detecting ovarian cancer | Gynaecological pathology | 10,296 patches, 174 patients + 58 cases for additional experiments | 2 | Mixed | Yes |
| Song[68] | 2013 | Basic Science Research Program, National Research Foundation of Korea, funded by the Ministry of Education, Science and Technology; INHA University Research Grant | Classifying types of pancreatic neoplasm | Hepatobiliary pathology | 11 WSIs, 400 patches | 1 | No | Unclear |
| Song[44] | 2020a | CAMS Innovation Fund for Medical Sciences; National Natural Science Foundation of China (NSFC); Tsinghua Initiative Research Programme. | Detecting colorectal adenomas | Gastrointestinal pathology | 579 WSIs | 3 | No | Yes |
| Song[93] | 2020b | National Natural Science Foundation of China; CAMS Innovation Fund for Medical Sciences; Medical Big Data and Artificial Intelligence Project of the Chinese PLA General Hospital; Tsinghua Initiative Research Program Grant; Beijing Hope Run Special Fund of Cancer Foundation of China. | Detecting gastric cancer | Gastrointestinal pathology | 8153 WSIs | 2 | No | Yes |
| Steinbuss[43] | 2020 | No Funders | Classify subtypes of gastritis | Gastrointestinal pathology | 1230 patches | 1 | No | Yes |
| Steiner[103] | 2018 | Google Brain Healthcare Technology Fellowship | Assist pathologist in detecting breast cancer metastases in lymph nodes | Breast pathology | 339 WSIs | 3 | Mixed | Yes |
| Sun[67] | 2020 | National Basic Research Program of China; Science and Technology Major Project of Hubei Province (Next-Generation AI | Detecting endometrial cancer; classifying endometrial diseases | Gynaecological pathology | 3502 patches | 1 | Mixed | Yes |
| | | Technologies); Medical Science and Technology projects of China | | | | | | |
| Swiderska-Chadaj[58] | 2020 | Philips Digital and Computational Pathology | Detecting prostate cancer | Uropathology | 717 WSIs | 3 | No | Yes |
| Syed[120] | 2021 | National Institute of Diabetes and Digestive and Kidney Diseases of the National Institutes of Health, Bill and Melinda Gates Foundation, University of Virginia Center for Engineering in Medicine, University of Virginia THRIV Scholar Career Development Award. | Detecting coeliac disease and environmental enteropathy | Gastrointestinal pathology | 461 WSIs | 3 | No | Yes |
| Syrykh[137] | 2020 | No funder declared | Detecting follicular lymphoma | Haematopathology | 491 WSIs (378 + 65 + 24 +24) | 2 | No | Yes |
| Tabibu[134] | 2019 | No funder declared | Detecting renal cancer and classifying subtype | Uropathology | 2105 WSIs | 1 | Yes | Yes |
| Tsuneki[42] | 2021 | No funders | Detecting poorly differentiated colorectal cancer | Gastrointestinal pathology | 2547 WSIs | 5 | No | Yes |
| Tsuneki[57] | 2022 | No funders | Detect prostate cancer | Uropathology | 3694 WSIs | 6 | Mixed | Yes |
| Tung[41] | 2022 | No funders declared | Detecting gastric cancer | Gastrointestinal pathology | 50 patients; 2750 image tiles. | 1 | Yes | Yes |
| Uegami[112] | 2022 | New Energy and Industrial Technology Development Organization (NEDO) | Detecting Usual Interstitial Pneumonia (UIP) | Cardiothoracic pathology | 715 WSIs + 181 WSIs pretraining set | 1 | No | Yes |
| Valkonen[104] | 2017 | 1. Academy of Finland<br>2. Tekes - The Finnish Funding Agency for Innovation<br>3. Cancer Society of Finland, Sigrid Juselius Foundation and Doctoral Programme of Computing and Electrical Engineering, Tampere University of Technology | Detecting breast cancer metastases in lymph nodes | Breast pathology | 270 WSIs | 1 | Yes | Unclear |
| Wang KS[48] | 2021 | 1. National Institutes of Health<br>2. Edward G. Schlieder Endowment and the Drs. W. C. Tsai and P. T. Kung Professorship in Biostatistics from Tulane University<br>3. National Key Research and Development Plan of China<br>4. National Natural Science Foundation of China<br>5. Jiangwang Educational Endowment.<br>6. Natural Science Foundation of Hunan Province | Detecting colorectal cancer | Gastrointestinal pathology | 14,680 WSIs | 14 | Mixed | Yes |
| Wang L[66] | 2020 | National Natural Science Foundation of China | Detect eyelid melanoma | Dermatopathology | 155 WSIs (83,126 patches) | 2 | No | Yes |
| Wang Q[50] | 2021 | National Natural Science Foundation of China, National KeyR&DProgram of China, KeyR&DProgram of Liaoning Province, Young and Middle-aged Talents Program of the National Civil Affairs Commission, Liaoning BaiQianWan Talents Program, University-Industry Collaborative Education Program. | Detecting breast cancer metastases in lymph nodes | Breast pathology | 529 WSIs | 2 | Yes | Yes |
| Wang S[39] | 2019 | Hong Kong Innovation and Technology Commission; Shenzhen Science and Technology Program. | Classification of gastric cancer and dysplasia | Gastrointestinal pathology | 608 WSIs | 1 | No | Yes |
| Wang X[65] | 2020 | Hong Kong Innovation and Technology Commission; National Natural Science Foundation of Chine; Shenzhen Science and Technology Program. | Classifying subtypes of lung cancer | Cardiothoracic pathology | 1439 WSIs (939 WSI internal, 500 WSI external) | 2 | Mixed | Yes |
| Wang C[40] | 2017 | National Natural Science Foundation of China | Detecting colorectal cancer | Gastrointestinal pathology | 10 WSIs (1000 150 x 150 pixel images) | 1 | Yes | Unclear |
| Wei[122] | 2020 | NIH; Geisel School of Medicine at Dartmouth; Norris Cotton Cancer Centre. | Classification of colorectal polyps | Gastrointestinal pathology | 746 WSIs | 2 | No | Yes |
| Wei[109] | 2019 | No funders declared | Classification of lung adenocarcinoma histological patterns | Cardiothoracic pathology | 422 WSIs | 1 | No | Yes |
| Wu[49] | 2020 | Information Technology for Cancer Research program and National Institutes of Health | Detecting breast cancer | Breast pathology | 240 cases | 1 | No | Unclear |
| Xu[140] | 2017 | Microsoft Research; Beijing National Science Foundation in China; Technology and Innovation Commission of Shenzhen in China; Beijing Young Talent Project in China; Fundamental Research Funds for the Central Universities of China from the State Key Laboratory of Software Development Environment in Beihang University in China. | Detecting & classifying brain cancer. Detecting colorectal cancer | Multiple | brain 141 images, colon 717 cropped regions | 2 | Mixed | Yes |
| Xu[38] | 2021 | Guangzhou Key Medical Discipline Construction Project Fund; Guangzhou Science and Technology Plan Project; Guangdong Provincial Science and Technology Plan Project. | Detecting colorectal cancer | Gastrointestinal pathology | 476 WSIs (263 + 218 -5 removed) | 2 | Mixed | Unclear |
| Yan[37] | 2022 | Science and Technology Innovation 2030-Key Project of China; Key-Area Research and Development Program of Guangdong Province, China. | Detecting colorectal cancer and colorectal polyps, detecting breast cancer lymph node metastases. | Multiple | NCT-CRC 100,000 patches. CAMELYON16 100,000 patches. In-house 20 patients. | 3 | Mixed | Unclear |
| Yang[110] | 2021 | National Key R&D Program of China; National Natural Science Foundation of China; Guangdong Natural Science Foundation; Support Scheme of Guangzhou for Leading Talents in Innovation and Entrepreneurship. | Classifying subtypes of lung cancer and other lung diseases | Cardiothoracic pathology | 1693 WSIs | 3 | Mixed | Yes |
| Yang[118] | 2022 | Ministry of Science and Technology (MOST), Taiwan | Detecting hepatocellular carcinoma | Hepatobiliary pathology | 46 WSIs | Unclear | Unclear | Yes |
| Yu[136] | 2020 | Schlager Family Award for Digital Health Innovations; Partners' Innovation Discovery Grant; Blavatnik Centre for Computational Biomedicine Award; Harvard Data Science Fellowship. | Detecting serous ovarian carcinoma & predicting tumour grade | Gynaecological pathology | 1375 WSIs | 1 | Yes | Yes |
| Yu[138] | 2019 | National Cancer Institute; National Institutes of Health; National Human Genome Research Institute; National Institutes of Health; Mobilize Centre, Stanford University; Harvard Data | Detecting lung cancer and classifying subtypes of lung cancer | Haematopathology | | 2 | Yes | Yes |
| | | Science Fellowship; Harvard Medical School Centre for Computational Biomedicine Award | | | | | | |
| Yu[139] | 2021 | No funders | Detecting T cell lymphomas & classifying T cell lymphoma subtypes | Haematopathology | 40 WSIs (1 per patient, 33 ROIs) | 17 | No | Yes |
| Zhang[64] | 2022 | Children's Cancer Fund of Dallas, the QuadW Foundation, the NIH grants NCI National Clinical Trials Network (NCTN) Operations Centre, NCTN SDC, Children's Oncology Group (COG) Biospecimen Bank, the Cancer Prevention and Research Institute of Texas. | Classifying subtypes of rhabodmyosarcoma | Soft tissue & bone pathology | 272 WSIs | 1 | Unclear | Yes |
| Zhao[63] | 2021 | Major Research Plan of the National Natural Science Foundation of China, the Shanghai Hospital Development Centre Clinical Science and Technology Innovation project, the National Key R&D Program of China and the National Natural Science Foundation of China. | Detecting lung cancer and classifying subtypes of lung cancer | Cardiothoracic pathology | 2125 WSIs | 1 | Yes | Yes |
| Zheng[111] | 2022 | National Institutes of Health, Johnson & Johnson Enterprise Innovation Inc., American Heart Association, Karen Toffler Charitable Trust, National Science Foundation. | Detecting lung cancer and classifying lung cancer subtypes | Cardiothoracic pathology | 4153 WSIs for train / validate / test + 665 WSIs used for earlier development | 3 | Yes | Yes |
| Zhou[127] | 2021 | Double-Class University project, the National Natural Science Foundation of China, and Postgraduate Research & Practice Innovation Program of Jiangsu Province | Detecting colorectal cancer | Gastrointestinal pathology | 1396 WSIs | 4 | Mixed | Yes |
| Zhu[56] | 2021 | US National Library of Medicine; US National Cancer Institute | Classify renal tumour subtypes | Uropathology | 1482 WSIs | 2 | Mixed | Yes |

*Given the varied language used to describe intended use, these were broadly categorised into detecting disease or classifying subtypes of disease for those relevant to this study.